\documentclass[pre,preprint]{revtex4}
\usepackage{amsmath}
\usepackage{epsfig}
\usepackage{color}
\begin{document}

\title{Percolation of particles on recursive lattices:\\ (II) the effect of size and shape disparities}
\author{Andrea Corsi\footnote{Present address: Dipartimento di Fisica, Universit\`a degli Studi di Milano, Milano, Italy}, P. D. Gujrati\footnote{Electronic address: pdg@physics.uakron.edu}}
\affiliation{The Department of Physics and The Department of Polymer Science, The University of Akron, Akron, Ohio, 44325, U.S.A.}
\date{\today}

\begin{abstract}

The preparation of many composites requires the intermixing of several macromolecular fluids along with the addition of rigid filler particles.
These fillers are usually polydisperse and there is an extensive experimental evidence that their size and shape profoundly affects the properties of the resulting material. In particular, it is generally found that the percolation threshold decreases as the size disparity between the different particles present in a system increases and that the threshold decreases with the aspect ratio of the particles. Here, a recursive approach that we have recently introduced is applied to the study of the percolation of particles of different sizes and shapes, without the presence of a polymer matrix, on a lattice in various phases including metastable states. In our approach, the original lattice is replaced by a recursive structure on which calculations are done exactly and interactions as well as size and shape disparities are easily taken into account. In the previous paper of this series, we have introduced the recursive approach and shown how correlations among particles of the same size can affect percolation. Before considering the complete system made of particles of various sizes and shapes embedded in a polymer matrix, in the third paper of the series, we describe here the properties of systems made of particles without any matrix. The approach appears to be extremely successful since it is able to capture most of the important features observed in experiments.

\end{abstract}

\maketitle

\section{Introduction}

\label{intro}

The fabrication of typical polymeric systems often requires the intermixing of several macromolecular fluids along with the addition of rigid filler particles. The most classical example of inclusion of rigid particles in a polymer matrix is represented by the addition of carbon black into a natural rubber matrix in order to improve its strength and processability. The original driving force for the incorporation of fillers into polymers was certainly economic since common polymers produced in high volume typically cost about ten times more than mineral fillers. In most cases, though, the cost argument cannot be sustained. The incorporation of fillers  into thermoplastic polymers through compounding, for example, is a costly process that almost always cancels out the economic advantage of introducing the filler.

The use of powdered materials of sizes ranging from nanometers to microns is not limited to polymer composites but is common in many branches of materials science. Fillers are carefully selected to enhance the performance of a matrix. Fillers have been used in order to improve mechanical \cite{Adams-93,Adams-Clay,Bigg-79,Chodak-99,Chodak-01,Ji-02,Kauly-96,Liang-99,Sharaf-01}, rheological \cite{Adams-93,Adams-Clay,Ghosh-00}, electrical \cite{Gokturk-93,Bigg-79,Chodak-99,Chodak-01,Lu-97,Mamunya-02,Michels-89,
Mikrajuddin-99, Narkis-00,Ota-97,Roldughin-00,Strumpler-99,Wang-92,Yi-99}, magnetic \cite{Gokturk-93,Fiske-97} and thermal \cite{Lu-97,Bigg-79,Dietsche-99,Hansen-75,Mamunya-02,Bigg-95} properties of the host material. The elastic modulus of a filled resin, as an example, results from a \emph{complex interplay} of the properties of the individual components. There is enough experimental evidence to conclude that the properties of the composite are affected by a large number of parameters: the size, shape and distribution of the reinforcing particles as well as the interactions between the particles and the polymeric matrix. Even without considering the interactions of the particles with the polymer matrix, it is clear from the experiments that the percolation threshold decreases as the size disparity between the different particles present in a system increases and that the threshold decreases with the aspect ratio of the particles present in the system \cite{Chodak-99,Kauly-96,Sharaf-01,Strumpler-99,Fiske-97}. In the case of particles that are not spherical in shape, another important parameter that affects the mechanical properties of the composite is represented by the degree of orientation of the particles with respect to the direction of an applied external stress state. 

Although most of the physical systems that are usually described through percolative models are polydisperse, most studies have described the percolation of equal sized components. Even though lattice models are simpler, the effect of polydispersity has been considered only in continuum percolation, mainly in the case of a system of particles represented by discs and spheres with a distribution of radii. In contrast to some initial claims that the average fractional volume covered by the percolating species at the threshold would be constant \cite{Kertesz-82}, results that have been recently published have shown that the volume fraction occupied by the percolating species at the threshold increases as the ratio between the size of the constituent species becomes larger and larger, more significantly so when the ratio itself is very large \cite{Meester-94}.

Many other models have been developed in order to describe the percolation process. Some of them involve modifications of the classical theory of rubber elasticity to account for additional cross-links produced by the fillers \cite{Kraus-65}. Models have also been developed based on a double network description \cite{Reichert-93} and on the replica formalism \cite{Heinrich-93}. 

Monte-Carlo lattice simulations of filled polymers are also very common and often based on the rotational isomeric-state model \cite{Kloczkowski-93,Kloczkowski-94,Sharaf-94,Sharaf-01}. The simulations are applied to both polymer chains attached at one end to the filler surface and to free chains. All these simulations assume that the filler particles occupy the sites of a regular cubic lattice and the polymer chains occupy the space between the particles. These simulations were successful in predicting the anisotropic reinforcing properties of prolate particles \cite{Sharaf-01}.

\section{Objective of the present research}

As explained before \cite{firstpaper}, a composite that contains nanoparticles is characterized by nanoscopic inhomogeneities that cannot be described satisfactorily in a continuum model since the averaging processes performed during the calculations are done on length scales much larger than the size of the nanoparticles. A lattice model is more promising and can be used to capture this nanoscopic inhomogeneities. Previous lattice model calculations of percolation generally did not take into account the possible difference between the size of filler particles (with very few exceptions like \cite{Amritkar-98}) and usually neglected interactions. Such calculations have additional very important limitations represented by the use of random mixing approximation, the incompressibility of the model, and the necessity for the monomers and voids to have the same size.

In our approach, we replace the original regular lattice by a recursive lattice (RL), which is built up from its smaller parts in a recursive fashion. The two infinitely large recursive structures that we have extensively used are the Bethe lattice and the Husimi lattice. The choice of the recursive lattice to be used is dictated by the model being investigated. 

Our theory is given by the solution on a recursive lattice. These tree structures allow us to capture only weak correlations and are consequently not suitable, for example, for carrying out the calculation of critical exponents. However, other kinds of recursive lattices that do not have a tree structure can be used to obtain non-classical exponents, as we have done in \cite{PDG-PRA-90}. However, as our interest here is not in critical exponents, we will restrict ourselves to the theory obtained on tree-like recursive lattices. This theory has been applied by Gujrati and coworkers to study and describe a wide range of polymer systems and it has  provided extensive insight in their phase separation, critical points, loop formation in tree polymer gel, theta states, compressibility effects, immiscibility loop, the Kauzmann paradox and the ideal glass transition \cite{PDG-Bowman-JCP-99, PDG-JCP1-98,PDG-JCP2-98,PDG-Chhajer-JCP-98,PDG-Corsi-PRL-01,PDG-Rane-Corsi-PRE-03,PDG-Corsi-PRE-03,Corsi-Masterthesis}. Another benefit of this approach has been its applicability to study the thermodynamics of a system in confined geometries \cite{mukesh_1,mukesh_2,mukesh_3,mukesh_4,mukesh_5}.

Here we study the possible percolation of filler particles of different sizes and shapes in the system. Although size disparity alone without any interactions has been argued to induce phase separation in this kind of system, recent rigorous calculations \cite{PDG-PRE-01} have proven that no phase separation in an athermal fully packed state of hard particle mixtures on a lattice is possible merely due to size disparity.

Our goal is to study the dependence of percolation on the size disparity and nature of the interactions between different filler particles. In the following paper \cite{thirdpaper}, we describe the effects of the presence of a polymer matrix on the percolation of these particles.

Even though conventionally, when a solution is described, the name $solvent$ is attributed to the abundant species, in the following we will always refer to the smallest component present in the system as the solvent even when that species is not the abundant one. This does not mean that different species must be chemically different, it is just a convention to make the description easier to understand.

We study a system of monomeric and square particles and compare its behavior with a system of monomeric and star particles on a square Husimi lattice to determine the effects of the change in shape of the larger particles on the properties of systems made of particles of different sizes. In this and the following publications, we will simply call the square Husimi lattice the Husimi lattice for brevity.

\section{Monomeric and square particles}

\subsection{Model}

The first system we investigate is the one formed by monomeric particles and \textit{square} particles on a Husimi lattice, see Figures \ref{Husimi} and \ref{Figure:square+solvent}. Figure \ref{Husimi} shows a portion of a Husimi lattice, including the labeling of its sites. The Husimi lattice is built in a recursive fashion starting from a central plaquette. The different size of the plaquettes that belong to different generations is just a matter of convention: one might consider all plaquettes to have the same size but then a lattice would become impossible to draw because of the crowding that one has as one moves away from the origin of the lattice. Figure \ref{Figure:square+solvent} shows one element of a Husimi lattice, that is a square plaquette.  As shown in the figure, the square particles that we are considering are particles in the shape of squares occupying one elemental square of the Husimi lattice and, therefore, its four lattice sites. As explained before \cite{firstpaper}, we define ``above'' and ``below'' relative to the origin of the lattice (above meaning moving away from the origin) and we distinguish the sites inside a plaquette of the Husimi lattice as a base site (the one closest to the origin of the lattice), two intermediate sites and a peak site, opposite to the base one.

In order to be able to describe this system, we introduce three possible states that describe the configuration of the $m$th level site which is the base site of the $m$th level square. This site can be in the $\text{S}$ state if it is occupied by a solvent molecule, in the $\text{A}$ state if a corner of a square particle occupies the site and the square is \textit{above} it (so that all the $(m+1)$th level sites are occupied by the remaining corners of the square) or in the $\text{B}$ state if a corner of a square
particle occupies the site and the square lies \textit{below} it, see Figure \ref{Figure:square+solvent}.

As we have seen in the first part \cite{firstpaper}, we can associate a Boltzmann weight $w$ with every nearest-neighbor contact between particles of different species present in the system. The weight $w$ is determined by the excess interaction energy $\varepsilon $\ as follows: 
\begin{equation}
w\equiv \exp \left( -\beta \varepsilon \right) .
\end{equation}%
We will set $\varepsilon $ =1 to set the temperature scale, as was also done in the first paper. The activity for the monomeric species particle, is $\eta $, related to its chemical potential $\mu $ through 
\begin{equation}
\eta \equiv \exp \left( \beta \mu \right) =w^{-\mu }.  \label{activity}
\end{equation}

The total partition function for this system of $N$ sites can be written as: 
\begin{equation}
Z\equiv \sum w^{N_{\mathrm{c}}}\eta ^{N_{\mathrm{s}}}  \label{PF}
\end{equation}%
where $N_{\text{c}}$ is the number of contacts between particles of two different species and $N_{\text{s}}$ is the number of solvent particles present in the lattice. From the partition function of the system we can obtain the thermodynamic potential or free energy per site $F$ by means of 
\begin{equation}
Z=e^{-NF/kT}  \label{Z_FreeEnergy}
\end{equation}%
In the following, instead of using the conventional expression of the free energy (per site), we will use the dimensionless free energy 
\begin{equation}
\omega =-\beta F.  \label{FreeEnergy}
\end{equation}%
The change in sign implies that whenever we will look for stable phases, we will have to look for the maximum of the free energy and not for its minimum. Taking into account, as it usually happens, that the sum in the expression for $Z$ over all sets of states is determined by the equilibrium state corresponding to the average densities per site $\phi _{\text{c}}$ and $\phi _{\text{s}}$, where $\phi _{\text{c}}$ is the density of contacts between solvent molecules and square particles and $\phi_{\text{s}}$ is the density of solvent molecules, and given the definition of the entropy of the system, we can write for the entropy per site 
\begin{equation}
s=\omega -\beta \mu \phi _{\text{s}}+\beta \phi _{\text{c}}.  \label{entropy}
\end{equation}%
All thermodynamic densities in this work will be defined per site unless noted otherwise. The procedure that we use to obtain the above two densities has been outlined in the previous paper \cite{firstpaper}: it is necessary to consider the contribution to the total partition function given by configurations in which the solvent is at the origin (to obtain the value of $\phi _{\text{s}}$) or by those in which the particle at the origin is in ``contact'' with one or more particles of a different species (to obtain the value of $\phi _{\text{c}}$).

\subsection{Recursion Relations}

When the $m$th level site is in the S state, there are eight possible configurations for the three sites at the $(m+1)$th level as shown in Figure \ref{Figure:square_solvent configurations}.

Following \cite{firstpaper}, we can write the recursion relation for $Z_{m}(\text{S})$, the partial partition function for the $m$th branch of the Husimi lattice given that the $m$th level site is occupied by a solvent molecule, as a function of the $Z_{m+1}$: 
\begin{align}
Z_{m}\left( \text{S}\right) & =\eta \lbrack Z_{m+1}\left( \text{S}\right)
Z_{m+1}\left( \text{S}\right) Z_{m+1}\left( \text{S}\right)  \notag \\
& +3w^{2}Z_{m+1}\left( \text{S}\right) Z_{m+1}\left( \text{S}\right)
Z_{m+1}\left( \text{A}\right)  \notag \\
& +(2w^{2}+w^{4})Z_{m+1}\left( \text{S}\right) Z_{m+1}\left( \text{A}\right)
Z_{m+1}\left( \text{A}\right)  \notag \\
& +w^{2}Z_{m+1}\left( \text{A}\right) Z_{m+1}\left( \text{A}\right)
Z_{m+1}\left( \text{A}\right) ].
\end{align}

If the the $m$th level site is in the B state, we can write a similar expression for $Z_{m}\left( B\right) $ since the possible configurations of the three sites at the $(m+1)$th level are the same as in the previous case: 
\begin{align}
Z_{m}\left( \text{B}\right) & =w^{2}Z_{m+1}\left( \text{S}\right)
Z_{m+1}\left( \text{S}\right) Z_{m+1}\left( \text{S}\right)  \notag \\
& +(2w^{2}+w^{4})Z_{m+1}\left( \text{S}\right) Z_{m+1}\left( \text{S}\right)
Z_{m+1}\left( \text{A}\right)  \notag \\
& +3w^{2}Z_{m+1}\left( \text{S}\right) Z_{m+1}\left( \text{A}\right)
Z_{m+1}\left( \text{A}\right)  \notag \\
& +Z_{m+1}\left( \text{A}\right) Z_{m+1}\left( \text{A}\right) Z_{m+1}\left( 
\text{A}\right) ,
\end{align}%
where the Boltzmann weight $w$ appears every time a solvent molecule is a nearest neighbor of a square particle. The possible configurations of the $(m+1)$th level sites when the base site is in the $\text{B}$ state are shown in Figure \ref{Figure:square_below configurations}. Finally, if the the $m$th level is in the $\text{A}$ state we can write: 
\begin{equation}
Z_{m}\left( \text{A}\right) =Z_{m+1}\left( \text{B}\right) Z_{m+1}\left( 
\text{B}\right) Z_{m+1}\left( \text{B}\right).
\end{equation}%
The single configuration that the system can assume in this particular case is shown in Figure \ref{Figure:square_above configurations}.

We introduce the following ratios: 
\begin{equation}
x_{m}\left( \text{S}\right) \equiv \frac{Z_{m}\left( \text{S}\right) }{B_{m}}%
,x_{m}\left( \text{A}\right) \equiv \frac{Z_{m}\left( \text{A}\right) }{B_{m}%
}
\end{equation}%
and 
\begin{equation}
x_{m}\left( \text{B}\right) \equiv \frac{Z_{m}\left( \text{B}\right) }{B_{m}}%
=1-x_{m}\left( \text{S}\right) -x_{m}\left( \text{A}\right) ,
\end{equation}%
where we have introduced 
\begin{equation}
B_{m}\equiv Z_{m}\left( \text{S}\right) +Z_{m}\left( \text{A}\right)
+Z_{m}\left( \text{B}\right) .  \label{Bmforsquares}
\end{equation}%
The ratios satisfy the following sum rule at every generation
\begin{equation}
x_{m}\left( \text{A}\right) +x_{m}\left( \text{B}\right) +x_{m}\left( \text{S%
}\right) \equiv 1.  \label{sum_rule}
\end{equation}%
We express $B_{m}$ in terms of the partial partition functions of the $(m+1)$th level for completeness: 
\begin{align}
B_{m}& =\eta \lbrack Z_{m+1}\left( \text{S}\right) Z_{m+1}\left( \text{S}%
\right) Z_{m+1}\left( \text{S}\right) +3w^{2}Z_{m+1}\left( \text{S}\right)
Z_{m+1}\left( \text{S}\right) Z_{m+1}\left( \text{A}\right)   \notag \\
& +(2w^{2}+w^{4})Z_{m+1}\left( \text{S}\right) Z_{m+1}\left( \text{A}\right)
Z_{m+1}\left( \text{A}\right) +w^{2}Z_{m+1}\left( \text{A}\right)
Z_{m+1}\left( \text{A}\right) Z_{m+1}\left( \text{A}\right) ]  \notag \\
& +w^{2}Z_{m+1}\left( \text{S}\right) Z_{m+1}\left( \text{S}\right)
Z_{m+1}\left( \text{S}\right) +(2w^{2}+w^{4})Z_{m+1}\left( \text{S}\right)
Z_{m+1}\left( \text{S}\right) Z_{m+1}\left( \text{A}\right)   \notag \\
& +3w^{2}Z_{m+1}\left( \text{S}\right) Z_{m+1}\left( \text{A}\right)
Z_{m+1}\left( \text{A}\right) +Z_{m+1}\left( \text{A}\right) Z_{m+1}\left( 
\text{A}\right) Z_{m+1}\left( \text{A}\right)   \notag \\
& +Z_{m+1}\left( \text{B}\right) Z_{m+1}\left( \text{B}\right) Z_{m+1}\left( 
\text{B}\right) ,
\end{align}%
which is easy to derive.

The choice of the normalization factor $B_{m}$ used to obtain these ratios from the partial partition functions is critical. We always try to keep these ratios finite so that the denominator cannot go to zero. In order to enforce this, we always use combinations such as that above: our lattice is always completely filled with either squares or solvent molecules present at any temperature. Thus, at any temperature, the terms $Z_{m}\left(\text{S}\right) $, $Z_{m}\left( \text{A}\right) $ or $Z_{m}\left( \text{B}\right) $ cannot all be zero and all the ratios are always finite. At the fixed point of the recursion relations we have: 
\begin{equation}
x_{m}\left( \text{S}\right) =x_{m+1}\left( \text{S}\right) =s,x_{m}\left( 
\text{A}\right) =x_{m+1}\left( \text{A}\right) =a,x_{m}\left( \text{B}%
\right) =x_{m+1}\left( \text{B}\right) =b.
\end{equation}

In general, we call such a fixed point solution, in which the ratios do not change from level to level, a 1-cycle solution. The reason for this nomenclature will become clear below. Using the fixed point solution, we can express 
\begin{equation}
B_{m}\equiv B_{m+1}^{3}Q,  \label{amplitude_relation}
\end{equation}%
where $B_{m+1}=Z_{m+1}\left( \text{S}\right) +Z_{m+1}\left( \text{A}\right)+Z_{m+1}\left( \text{B}\right) $ according to equation \ref{Bmforsquares}, and where we have introduced a polynomial $Q$, which at the fix-point takes the value 
\begin{equation}
Q=(\eta +w^{2})s^{3}+(3\eta w^{2}+2w^{2}+w^{4})s^{2}a+(3w^{2}+2\eta
w^{2}+\eta w^{4})sa^{2}+(1+w^{2})a^{3}+b^{3}.  \label{amplitude_polynomial}
\end{equation}%
Using this polynomial and the recursion relations written above, we can write at the fix-point the following system of equations to be solved: 
\begin{subequations}
\begin{align}
sQ& =\eta \lbrack s^{3}+3w^{2}s^{2}a+(2w^{2}+w^{4})sa^{2}+w^{2}a^{3}],
\label{RR1} \\
aQ& =b^{3},  \label{RR2} \\
bQ& =w^{2}s^{3}+(2w^{2}+w^{4})s^{2}a+3w^{2}sa^{2}+a^{3}.  \label{RR3}
\end{align}

\subsection{Phase Diagram and Percolation}

\subsubsection{Densities}

In order to obtain the densities of the two species, we must consider the total partition function of our system. The total partition function of the system at the $(m=0)$th level can be written considering all the possible configurations of the $(m=0)$th level site \cite{PDG-PRL-95}. The entire lattice is obtained by joining two $(m=0)$th level branches $\mathcal{C}_{0}$; see \cite{firstpaper} for definition. In order to obtain the total partition function, we must consider all the configurations that the system can assume in the two branches that meet at the origin of the lattice.

There are only three possible configurations that the system can assume at the origin, as shown in Figure \ref{Figure:square+solvent}. The site at the origin can be occupied by either a solvent molecule or the corner of a square particle. In the case of a square particle, we have to take into account the fact that the square can be either above or below the origin. The total partition function of the system can then be written as  
\end{subequations}
\begin{equation}
Z_{0}=\frac{Z_{0}(\text{S})Z_{0}(\text{S})}{\eta }+2Z_{0}(\text{A})Z_{0}(%
\text{B}).  \label{Equation: Sq+Solv Total PF}
\end{equation}
The first term represents the contribution of the configuration with the solvent molecule at the origin, while the second one represents the contribution of the configurations with the square at the origin. As explained before \cite{firstpaper}, the factor $\eta $ in the denominator of the first term is necessary in order not to over-count the solvent activity. The factor two in the last term is related to the possibility of having the square on either side of the origin.

The form of this contribution is easily understood if we realize that every configuration that appears as an $\text{A}$ state looking at it from one side of the origin, appears to be in the $\text{B}$ state if considered from the opposite side.

The solvent density is then defined as the ratio between that part of the total partition function containing configurations in which the origin is occupied by a solvent molecule and the total partition function of the system.

In this particular case, we can express the density of the solvent particles on the lattice as: 
\begin{equation}
\phi _{\text{s}}=\frac{Z_{0}\left( \text{S}\right) Z_{0}\left( \text{S}%
\right) /\eta }{Z_{0}}.  \label{solvent_density}
\end{equation}%
Similarly, the density of sites covered by the square particles is defined as 
\begin{equation}
\phi _{\text{sq}}=\frac{2Z_{0}\left( \text{A}\right) Z_{0}\left( \text{B}%
\right) }{Z_{0}}.  \label{square_density0}
\end{equation}%
This is, by the definition, the mass density of squares. It is obvious that $\phi _{\text{sq}}+\phi _{\text{s}}\equiv 1.$ Since each square has four corners (and occupies four sites), the number density of squares is 
\begin{equation}
\phi _{\text{sq,n}}\equiv \phi _{\text{sq}}/4.  \label{square_density1}
\end{equation}

\subsubsection{Free Energy per Site}

The total partition function can be used to obtain the thermodynamics of the system. It is clear that $Z_{0}$ is the total partition function of the system obtained by joining two branches $\mathcal{C}_{0}$ together at the origin. In the thermodynamic limit, $Z_{0}$ becomes unbounded and care must be exercised. For this purpose, we need to consider the free energy per site. This we accomplish as follows. Let us imagine taking away from the lattice the two squares that meet at the origin. This leaves behind six different branches $C_{1}$. We connect these branches to form three smaller but identical cacti, the partition function of each of which is denoted by $Z_{1}$; the latter can be written in a form that is identical to that of equation \ref{Equation: Sq+Solv Total PF},
except that the index 0 of each partial partition function is replaced by 1.

The difference between the free energy of the complete lattice and that of the three reduced lattices is just the free energy corresponding to a pair of squares (which contain 4 sites of the lattice) so that, following Gujrati \cite{PDG-PRL-95}, we can write the adimensional free energy per site without the conventional minus sign as 
\begin{equation}
\omega =\frac{1}{4}(\omega (\text{complete lattice})-3\omega (\text{reduced
lattice}))=\frac{1}{4}\ln \left( \frac{Z_{0}}{Z_{1}^{3}}\right) .
\end{equation}

It is possible to write 
\begin{equation}
Z_{0}=B_{0}^{2}Q_{2},Z_{1}=B_{1}^{2}Q_{2},
\end{equation}%
where $B_{m}$ is defined above and $Q_{2}$ is, at the fixed point, the polynomial 
\begin{equation}
Q_{2}=\frac{s^{2}}{\eta }+2ab.  \label{PF_polynomial}
\end{equation}%
Since $B_{0}=B_{1}^{3}Q,$ see (\ref{amplitude_relation}), the free energy per site can be written as 
\begin{equation}
\omega =\frac{1}{2}\ln \left( \frac{Q}{Q_{2}}\right) .  \label{FreeEnergy0}
\end{equation}

The calculations are done in the grand canonical ensemble and are carried out at constant chemical potentials. Hence, the above free energy represents $\beta P_{\mathrm{sq}}v_{0}$\ where $\beta $ is the inverse temperature, $v_{0}$ represents the volume of the unit cell of the lattice and $P_{\mathrm{sq}}$ is the osmotic pressure \cite{PDG-JCP3-98} across a membrane permeable to the square particles. The osmotic pressure $P_{\mathrm{s}}$ across a membrane permeable to the solvent particles is given by $\beta P_{\mathrm{s}
}v_{0}\equiv \beta P_{\mathrm{sq}}v_{0}-\ln \eta.$ If the solvents represent voids, then $P_{\mathrm{s}}$ represents the conventional pressure of the lattice system \cite{PDG-JCP3-98}. The approach that we use is the following: for every value of $w$ ($T$) we solve the system of recursive equations that we have derived. Then we use the fixed point values of $a$, $b $ and $s$ to calculate the value of the free energy at that $w$.

In order to determine which phase is the stable one at some temperature, we must find the free energy of all the possible phases of the system as a function of $w$.

\subsubsection{$\protect\mu >0$}

Let us first consider the case when $\mu $ is positive. The ground state of the lattice at zero temperature is represented by a pure solvent phase in this case: all the lattice sites are occupied by monomeric species. As we start increasing the temperature, the density of squares increases. In this case, the solvent is percolating at every temperature. Thus, its percolation is of no interest. 

For positive values of $\mu $, only one solution of the set of recursive equations exists and consequently only one phase is present and, therefore, no phase transition is observed. From what was said earlier, this phase is easily seen to be a mixture of solvent particles and squares, which contains only solvent particles at absolute zero. As no phase transition occurs, we need to only investigate is the percolation of squares to which we now turn. 

To study the percolation of the square particles, we introduce as before the probability $R_{m}(\text{A})\leq 1$ that a site in an $\text{A}$ state at the $m$th generation is connected to a finite cluster of squares at higher generations as well as a probability $R_{m}(\text{B})\leq 1$ that a site in a $\text{B}$ state at the $m$th generation is connected to a finite cluster of squares at higher generations. Then $Z_{m}(\alpha )R_{m}(\alpha )$ denotes the contribution to the partial partition function of the branch $\mathcal{C}_{m}$ due to all those configurations in which the site at the $m$th generation is in the state $\alpha $ and is connected to a finite cluster of squares at higher generations. If we divide $Z_{m}(\alpha )R_{m}(\alpha )$ by $Z_{m}(\alpha )$ we obtain a recursion relation for the $R_{m}(\alpha )$. At the fixed-point solution, each one of these $R_{m}(\alpha )$ approaches its fixed-point value given by the solution of an equation of the form 
\begin{equation}
R(\alpha )=\rho \lbrack \left\{ R(\beta )\right\} ],
\label{percolation_relation}
\end{equation}%
where $\left\{ R(\beta )\right\} $ represents the set formed by the complete set of states $\beta $. By considering the configurations shown in Figures \ref{Figure:square_below configurations} and \ref{Figure:square_above configurations}, and following an approach that is identical to that used to obtain the partial partition functions, we can write: 
\begin{align}
R(\text{A})& =R(\text{B})^{3}, \\
R(\text{B})& =\frac{w^{2}s^{3}+(2w^{2}R(\text{A})+w^{4})s^{2}a+3w^{2}R(\text{%
A})^{2}sa^{2}+R(\text{A})^{3}a^{3}}{%
w^{2}s^{3}+(2w^{2}+w^{4})s^{2}a+3w^{2}sa^{2}+a^{3}}.
\end{align}

This system of equations does not depend explicitly on the chemical potential $\mu $ but it depends on it implicitly. In fact, for every value of $T$, and hence of $w$, a different chemical potential corresponds to different values of $s$, $a$ and $b$ and consequently a different value for the $R(\alpha )^{\prime }$s.

Since $R(\alpha )$ represents the probability of the (occupied) site at the origin of the lattice being connected to a finite cluster of squares on one half of the lattice and since two half lattices are joined together at the origin to form the complete lattice, we define 
\begin{equation}
p_{\text{sq}}=1-R(\text{A})R(\text{B})
\end{equation}%
as the percolation probability for the squares. This represents the probability of the occupied site at the origin being connected to an infinite cluster of squares on both sides of the origin. The behavior of $p_{\text{sq}}$ as a function of $\phi _{\text{sq}}$ for different (positive) values of $\mu $, is shown in Figure \ref{Figure:husimi_sq_solv_pos_mu_pvsphi}. As expected, the percolation
threshold increases as the chemical potential increases. The athermal case, when the two species do not have any interactions, corresponds to $\mu=+\infty ,\varepsilon \rightarrow 0$ such that $\mu \varepsilon $ $>0$ is kept constant.

As before, we can study the percolation process as a function of the temperature. Figure \ref{Figure:husimi_sq_solv_pos_mu_pvst} shows the percolation probability as a function of the temperature for the same chemical potential values considered in Figure \ref{Figure:husimi_sq_solv_pos_mu_pvsphi}.

Figure \ref{Figure:husimi_sq_solv_pos_mu_phicvsmu} shows how the percolation threshold depends on the value of the chemical potential while Figure \ref{Figure:husimi_sq_solv_pos_mu_tcvsmu} shows the dependence of the critical temperature. As expected, an increase in the chemical potential for the smaller particles makes it more and more difficult for the bigger particles to percolate. This explains the increase in the critical temperature and the percolation threshold as $\mu $ increases. As before, the limiting value for the percolation threshold as $\mu $ goes to infinity is the athermal value while the critical temperature grows unbounded.

\subsubsection{$\protect\mu <0$}

If negative values of $\mu $ are considered, the problem becomes more interesting and slightly more complicated. A negative value for $\mu $ means that, according to thermodynamics, the system will be stable at low temperature if no solvent is present in the lattice and all the sites are covered by square particles. The very nature of the square particles, in particular the fact that they occupy more than one single site on the lattice, makes the solution of the problem more challenging. What can be expected at low temperature, or at very low concentrations of solvent molecules is an ordered structure of squares. The difference between the nature of this phase and that of a pure solvent phase must be appreciated. 
The \textit{order} present in the case of a pure solvent phase is due to the regular nature of the lattice (the periodic nature in the
case of a regular lattice), and has nothing to do with the nature of  the interactions between the particles that are present in the system. Instead, when a phase that is very rich in square particles is present, the \textit{order} of the structure is associated with the nature of the particles and is not only due to the lattice itself.

In the present case, we expect to have an ordered phase at low temperature that eventually disorders at high temperature, when large fractions of both solvent and square particles are present in the system. Thus, we expect to have a phase transition between an ordered phase, stable at low temperatures, and a disordered phase, stable at higher temperatures. The empty circles in Figure \ref{Figure:husimi_sq_solv_neg_mu_omegavst} represent the free energy of the system as obtained by using the above method of 1-cycle calculation in the case of $\mu =-1$. This \textit{cannot} be the free energy corresponding to the stable phase at every temperature, since it becomes negative. The adimensional free energy $\omega $ is the logarithm of the partition function. If we are properly describing the ground state of our system, the partition function at $T=0$ reduces to 2, since the only configuration that contributes is the lattice fully packed by square particles. On a square lattice, there are only two such configurations corresponding to the checkerboard packing of the lattice. Thus, the partition function at any positive temperatures contains this term besides other positive terms. 
Hence, $\omega $ for a stable phase \textit{cannot} be negative. This means that the phase described by the 1-cycle solution is not describing the stable phase at low temperatures because of the phase transition. It is, therefore, necessary to introduce a different description of our configurations on the lattice in order to be able to capture the phase that is stable at low temperatures. This is a problem that we have already encountered in previous studies \cite{PDG-Corsi-PRL-01,PDG-Corsi-PRE-03} and it requires a different fixed point solution than the conventional 1-cycle solution used for dealing with disordered phases.

The presence of large square particles introduces a \emph{correlation} between sites that are not nearest neighbors. In order to describe these correlations, let us first consider a system made of species that are all monomeric in size. If the base site of a particular $m$th level plaquette is occupied by a particle of one species, then the middle sites and the peak site of the same plaquette can be occupied in principle by a particle belonging to any species. In the presence of larger particles that occupy multiple sites of the lattice, this is not true anymore. A first example of this has been presented above when we were looking at the possible configurations of the system when the site of interest is in the $\text{A}$ state, as shown in Figure \ref{Figure:square_above configurations}. In this  case the two middle sites and the peak site do not have any freedom and can only be in the $\text{B}$ state. The recursion relation for $Z_{m}(\text{A})$ in fact contains one single term. But the correlations introduced by the presence of these large particles affect even larger portions of the lattice. If we move one level beyond to the $(m+1)$th plaquette, we immediately understand that the presence of the square particle inside the $m $th plaquette affects the state of the sites in the $(m+1)$th plaquette as well. The middle sites and the peak site in the $(m+1)$th plaquette cannot be in the $\text{B}$ state because that would be incompatible with the state of the sites one level below. From this discussion, it should be clear how the introduction of non-monomeric species induces correlations that would have not been present if we just had monomeric species.

From the point of view of the recursion relations, it is necessary to look for a different kind of fixed point solution. When using our recursive approach, we always start from some initial guesses for the ratios that describe the possible states of any site of the lattice. These initial guesses can be thought of as representing some particular boundary conditions at the surface of the system. Then, by using the appropriate recursion relations for the problem under investigation, we descend the lattice and we obtain the values of the ratios at the $m$th generation of the lattice as a function of the values at the $(m+1)$th generation. When the difference between the values of the ratios at two consecutive levels is lower than a preset tolerance, usually set equal to  10$^{-15}$, it is assumed that the fixed point has been reached and that the values obtained for the ratios describe the behavior in the bulk of the system. The values of the ratios at the fixed point can be used to obtain the free energy of the system and the entire thermodynamics as explained in the previous section. Different initial guesses are chosen in order to investigate if different fixed points, and consequently different phases, are present in the system for any temperature. This method, as explained above, is what we refer to as the \textit{one-cycle method}. This explains the meaning of the label for the lower free energy curve in Figure \ref{Figure:husimi_sq_solv_neg_mu_omegavst}. 

This approach never captures the ordered phase of square particles like the one that is expected at low temperatures in the case of negative chemical potentials. To obtain the appropriate description for the ordered phase at low temperatures, we start with the state at absolute zero, where all the lattice sites are occupied by the square particles. If the base site (index $m$) of a plaquette is in the $\text{A}$ state then the middle sites (index $m+1$) and the peak site (index $m+1$) of that plaquette are in the $\text{B}$ state. The middle and peak sites (index $m+2$) in the following plaquette are in the $\text{A}$ state and so on. It is possible to
notice a periodic structure in which consecutive generations on the lattice are alternatively in the $\text{A}$ and $\text{B}$ states. The one-cycle approach to the solution of the recursion relations will not succeed in describing this ordered phase.  What we have is this: if at one generation $m$ of the lattice $a=0$, for example, then at the following generation $m+1,$ $a=1$ followed by $a=0$ and so on so forth. In this case, the difference $a_{m}-a_{m+1}$ never goes to zero, but the difference $a_{m}-a_{m+2}$ does. This provides us with the clue to describing the low temperature phase ordered phase where the system is mostly filled with square particles as follows. We need to use the \textit{two-cycle } description for the ordered phase at low temperatures. In this case what is checked is the difference of the values of the ratios at two levels that are one generation apart on the lattice. Whenever $a_{m}-a_{m+2}$ (as well as $b_{m}-b_{m+2}$ and $s_{m}-s_{m+2}$) but not $a_{m}-a_{m+1}$ (as well as $b_{m}-b_{m+1}$ and $s_{m}-s_{m+1}$) is less than the desired tolerance, we assume the fixed point of the solutions has been reached. The system of equations for this kind of fix-point is the same as derived above and the equations are used to calculate the quantities of interest needed to describe the thermodynamics of the problem as well as the percolation probabilities for the system. 

It should be stressed that the two-cycle fix-point contains the one-cycle fix-point as a special case when the quantities at two successive generations become the same. Of course one might also obtain higher order cycles in which quantities such as $a_{m}-a_{m+n}$ with $n>2$ go to zero as the fixed point is approached. 
For an example of such a situation, we refer the reader to the following section. In the case of the system under investigation we have tried to look at three-, four- and five-cycle solutions without getting any physically meaningful results. But we do have to point out that, at least in principle, one can not rule out the presence of higher order cycles without actually looking for them, even though they are hard to envision. One could expect a 4-cycle structure in the present case if there were attractive interactions ($w>1$) such as specific interactions like hydrogen bonding between the solvent and the solute. But we do not consider these interactions. However, in general, the available cycle structures would emerge from the same set of recursive relations by starting from all possible initial guesses, and we have searched for different ones but we have been able to find just the two possible solutions mentioned in this section.

By using this second method, it is possible to obtain a second fixed point solution. The corresponding free energy curve is the upper one represented by the circles Figure \ref{Figure:husimi_sq_solv_neg_mu_omegavst}. The solutions obtained with the two different schemes coincide at high temperatures, but are different at low temperatures. The temperature at which the ordered phase appears and becomes the stable solution is, in the case of $\mu =-1$, $T_{\mathrm{OD}}\simeq 15.7$, indicated by the dotted line in the figure. We use the subscript OD to remind that this transition is a phase transition between an ordered and a disordered phase. It is possible to observe that the stable phase at every temperature has a positive free energy, as expected. Since $\omega $ lacks the conventional
minus sign, the stable phase is the one that has the \textit{maximum free energy}.

The entropy corresponding to the free energy shown in Figure \ref{Figure:husimi_sq_solv_neg_mu_omegavst} is shown in Figure \ref{Figure:husimi_sq_solv_neg_mu_entropyvst}.

As expected, below $T_{\text{OD}}$ the entropy of the ordered phase is lower than the entropy of the metastable continuation of the disordered phase. The entropy of the metastable phase, though, drops much more rapidly, and goes to zero at a finite temperature. This temperature is the Kauzmann temperature of the system, corresponding to the Kauzmann catastrophe \cite{Kauzmann-48}. This result is very interesting since until recently the Kauzmann catastrophe had been observed theoretically only in polymeric systems \cite{PDG-Corsi-PRL-01,PDG-Corsi-PRE-03}, but never in system made of small molecules. Recently Semerianov \cite{Semerianov-03} has shown the presence of a Kauzmann catastrophe in a dimer model.

The results for the percolation probability as a function of the temperature are shown in Figure \ref{Figure:husimi_sq_solv_neg_mu_pvst}. It is obvious that the strength of the percolation process is very different for the two phases. The formation of a percolating cluster of solvent molecules is much easier in the disordered phase than in the ordered one. There is a wide temperature range, between $\sim 8.42$ and $\sim 2.25$ in the case $\mu =-1$, in which there is at least one percolating cluster in the disordered phase but no percolation occurs in the ordered phase. The region below $2.25$, limited by the dotted line in Figure \ref{Figure:husimi_sq_solv_neg_mu_pvst} does not represent physical states of the disordered phase because it is below the Kauzmann temperature and it corresponds to a negative entropy.

The percolation probability is plotted as a function of the solvent density in Figure \ref{Figure:husimi_sq_solv_neg_mu_pvsphis}. This figure shows how the percolation threshold for the two phases is very different. In particular the solvent density must be at least $\sim 0.34$ in order to have percolation in the ordered phase while it is much lower in the metastable prosecution of the disordered phase. While lowering the temperature, if the phase transition is avoided the system is percolating all the way down to the Kauzmann temperature.

Figures \ref{Figure:husimi_sq_solv_neg_mu_tcvsmu}\ and \ref{Figure:husimi_sq_solv_neg_mu_phicvsmu}\ summarize the results obtained as
the value of the solvent chemical potential is changed. As expected, an increase in the magnitude of the chemical potential makes it more and more difficult for the solvent particles to percolate because such an increase makes it more and more favorable for the lattice to be covered by square particles reducing as much as possible the number of solvent particles present in the system, at least at low temperature.

If we consider the athermal limit in which there is no interaction between the two species of particles, we obtain the value $\phi _{s}\simeq 0.23$ and $\phi _{sq}\simeq 0.255$. These results are very interesting for two reasons. First of all, we observe that the percolation threshold in this athermal case is larger for the square particles than for the solvent particles. This is consistent with most experimental findings that show how the percolation threshold increases with the size of the particles. Also, the percolation threshold is lower for both species with respect to the value obtained for same size species in other studies, $\phi _{s}\simeq 0.38$. This behavior is also consistent with most of the experimental findings that the presence of size disparity lowers the percolation threshold of all the species present in the system.

\section{Monomeric and 5-site star particles }

\subsection{Model}

The next system that we have analyzed is the one formed by monomeric particles and \textit{star }particles on a Husimi lattice, see Figure \ref{Figure:star+solvent}; compare it with Figure \ref{Figure:square_solvent configurations}. We only consider the smallest possible stars, each of which occupies five sites of the lattice, as compared to the four sites occupied by the square particles considered above, and always have their core (the center) on a corner of a square plaquette of the lattice. Consequently, the star particles always occupy sites belonging to two different and adjacent plaquettes of the lattice. The introduction of the star shaped particles introduces correlations among sites of the lattice that have a range that is longer than the one characteristic of the squares. This makes the problem a bit more challenging. In order to be able to describe this system we introduced four possible states that describe the configuration of the $m$th level site of the lattice. The base site in the $m$th level square can be in an $\text{A}$ state if an end point of a star particle occupies the site and the core of the star is \textit{above} it, a $\text{B}$ state if an end point of the star occupies the site and the core of the star lies \textit{below} it or a $\text{C}$ state if the core of a star occupies the site, see Figure \ref{Figure:star+solvent}. It is important to point out that the $\text{A}$ and $\text{B}$ states describe all configurations in which an end point is at the site of interest with a core above or below, respectively. So parts (a) and (b) of Figure \ref{Figure:star+solvent} only show one of the two possible A and B configurations each. The star could be also on the other side of the plaquette as well. The site is in the $S$ state if it is occupied by a solvent molecule. As before, we always refer to \textit{above} and \textit{ below} with respect to the origin of the lattice: \textit{above} means farther away from the origin of the lattice while \textit{below} means closer to the origin.

Following what has been done before in the case of square particles, we can associate a Boltzmann weight $w$ with every contact between particles of different nature present in the system. Like before, the weight $w$ is determined by the excess interaction energy $\varepsilon $ as $w=\exp \left( -\beta \varepsilon \right) $. The activity for the solvent, $\eta $, is obtained as a function of its chemical potential, $\mu $, through $\eta =\exp \left( \beta \mu \right) =w^{-\mu }$ as before; see (\ref{activity}).

Similar to the case of monomeric and square particles, the total partition function for this system can be written as (\ref{PF}). The entropy per site is similarly given by (\ref{entropy}).

\subsection{Recursion Relations}

When a site at the $m$th level is in the $S$ state, there are nine possible configurations for the three sites at the $(m+1)$th level. These nine configurations are shown in Figure \ref{Figure:star_solvent configurations}. In the figure, the stars are always shown on one side but it is important to remember that, as explained above, they could be on either side of the plaquette. The configurations are very similar to those obtained in the previous section for the percolation of monomeric and square particles with the important difference that in the case of the stars it is possible to have the configuration in which the core of a star is at the peak site of the $(m+1)$th plaquette of the lattice with two arms of the stars being inside such plaquette (configuration (i) in the figure). In all configurations, the stars occupies sites of the lattice belonging to \emph{two} different plaquettes.

By using the same considerations as in the previous section, we can construct the recursion relation for $Z_{m}(\text{S})$, the partial partition function for the $m$th branch of the Husimi lattice given that the $m$th level site is occupied by a solvent molecule as: 
\begin{align}
Z_{m}\left( \text{S}\right) & =\eta \lbrack Z_{m+1}\left( \text{S}\right)
Z_{m+1}\left( \text{S}\right) Z_{m+1}\left( \text{S}\right)
+3w^{2}Z_{m+1}\left( \text{S}\right) Z_{m+1}\left( \text{S}\right)
Z_{m+1}\left( \text{A}\right)   \notag \\
& +(2w^{2}+w^{4})Z_{m+1}\left( \text{S}\right) Z_{m+1}\left( \text{A}\right)
Z_{m+1}\left( \text{A}\right) +w^{2}Z_{m+1}\left( \text{A}\right)
Z_{m+1}\left( \text{A}\right) Z_{m+1}\left( \text{A}\right)   \notag \\
& +w^{2}Z_{m+1}\left( \text{C}\right) Z_{m+1}\left( \text{B}\right)
Z_{m+1}\left( \text{B}\right) ],
\end{align}%
where $\eta $ is the activity.

If the site at the $m$th level is in the $\text{A}$ state, we can write: 
\begin{equation}
Z_{m}\left( \text{A}\right) =2Z_{m+1}\left( \text{C}\right) Z_{m+1}\left( 
\text{B}\right) Z_{m+1}\left( \text{A}\right) +2w^{2}Z_{m+1}\left( \text{C}%
\right) Z_{m+1}\left( \text{B}\right) Z_{m+1}\left( \text{S}\right) .
\end{equation}

The configurations that the system can assume in this particular case are shown in Figure \ref{Figure:star_above configurations}.

If the site at the $m$th level is in the $\text{B}$ state, we can write an expression that is very similar to the one written above for the $S$ state since the possible configurations of the three sites at the $(m+1)$th level are the same as in that case and only the weights have to be properly changed. Then by properly taking into account all the nearest neighbors interactions between particles of different species, we finally have: 
\begin{align}
Z_{m}\left( \text{B}\right) & =w^{2}Z_{m+1}\left( \text{S}\right)
Z_{m+1}\left( \text{S}\right) Z_{m+1}\left( \text{S}\right)
+(2w^{2}+w^{4})Z_{m+1}\left( \text{S}\right) Z_{m+1}\left( \text{S}\right)
Z_{m+1}\left( \text{A}\right)   \notag \\
& +3w^{2}Z_{m+1}\left( \text{S}\right) Z_{m+1}\left( \text{A}\right)
Z_{m+1}\left( \text{A}\right) +Z_{m+1}\left( \text{A}\right) Z_{m+1}\left( 
\text{A}\right) Z_{m+1}\left( \text{A}\right)  \\
& +Z_{m+1}\left( \text{C}\right) Z_{m+1}\left( \text{B}\right) Z_{m+1}\left( 
\text{B}\right) ,
\end{align}%
where the Boltzmann weight $w$ appears every time a solvent molecule is a nearest neighbor of a star particle. The possible configurations of the $(m+1)$th level sites are shown in Figure \ref{Figure:star_below configurations}. Finally, if the site at the $m$th level is in the $\text{C}$ state, we have: 
\begin{equation}
Z_{m}\left( \text{C}\right) =Z_{m+1}\left( \text{B}\right) Z_{m+1}\left( 
\text{B}\right) Z_{m+1}\left( \text{A}\right) +w^{2}Z_{m+1}\left( \text{B}%
\right) Z_{m+1}\left( \text{B}\right) Z_{m+1}\left( \text{S}\right) .
\end{equation}%
The configurations that the system can assume in this particular case are shown in Figure \ref{Figure:star_core configurations}.

Following the approach introduced in the previous section, we introduce the following ratios: 
\begin{equation}
x_{m}\left( \text{S}\right) =\frac{Z_{m}\left( \text{S}\right) }{B_{m}}%
,x_{m}\left( \text{A}\right) =\frac{Z_{m}\left( \text{A}\right) }{B_{m}}%
,x_{m}\left( \text{B}\right) =\frac{Z_{m}\left( \text{B}\right) }{B_{m}}
\end{equation}%
\begin{equation}
x_{m}\left( \text{C}\right) =\frac{Z_{m}\left( \text{C}\right) }{B_{m}}%
=1-x_{m}\left( \text{S}\right) -x_{m}\left( \text{A}\right) -x_{m}\left( 
\text{B}\right) .
\end{equation}%
where we have introduced 
\begin{equation}
B_{m}=Z_{m}\left( \text{S}\right) +Z_{m}\left( \text{A}\right) +Z_{m}\left( 
\text{B}\right) +Z_{m}\left( \text{C}\right) .  \label{Bmforstars}
\end{equation}%
The ratios satisfy the following sum rule at every generation
\begin{equation}
x_{m}\left( \text{A}\right) +x_{m}\left( \text{B}\right) +x_{m}\left( \text{C%
}\right) +x_{m}\left( \text{S}\right) \equiv 1;
\end{equation}
compare with (\ref{sum_rule}). 

At the (1-cycle) fixed point of the recursion relations, we have: 
\begin{align}
x_{m}\left( \text{S}\right) & =x_{m+1}\left( \text{S}\right) =s,  \notag \\
x_{m}\left( \text{A}\right) & =x_{m+1}\left( \text{A}\right) =a,  \notag \\
x_{m}\left( \text{B}\right) & =x_{m+1}\left( \text{B}\right) =b,  \notag \\
x_{m}\left( \text{C}\right) & =x_{m+1}\left( \text{C}\right) =c,
\end{align}%
Then, at this fix-point, we can introduce a polynomial $Q,$ as done in (\ref{amplitude_relation}), which is given by 
\begin{eqnarray}
Q &=&(\eta +w^{2})s^{3}+(3\eta w^{2}+2w^{2}+w^{4})s^{2}a+(3w^{2}+2\eta
w^{2}+\eta w^{4})sa^{2}  \notag \\
&&+(1+w^{2})a^{3}+b^{3}+(\eta w^{2}+1)cb^{2}+2(a+w^{2}s)bc+(a+w^{2}s)b^{2}.
\end{eqnarray}%
Using this polynomial and the recursion relations written above, we can write 
\begin{align}
sQ& =\eta \lbrack
s^{3}+3w^{2}s^{2}a+(2w^{2}+w^{4})sa^{2}+w^{2}a^{3}+w^{2}cb^{2}], \\
aQ& =2(a+w^{2}s)bc, \\
bQ& =w^{2}s^{3}+(2w^{2}+w^{4})s^{2}a+3w^{2}sa^{2}+a^{3}+cb^{2}, \\
cQ& =(a+w^{2}s)b^{2}.
\end{align}

\subsection{Phase Diagram and Percolation}

\subsubsection{Densities and Free Energy per Site}

In order to obtain the densities of the two species, we must consider the total partition function of our system.

There are only three possible configurations that the system can assume at the origin, as shown in Figure \ref{Figure:star+solvent}. The site at the origin can be occupied by either a solvent molecule, or the end point of a star or the core of a star. In the case of a star particle, we have to take into account the fact that the end point of the star can be either above or below the origin. The total partition function of the system can then be written as 
\begin{equation}
Z_{0}=\frac{Z_{0}\left( \text{S}\right) Z_{0}\left( \text{S}\right) }{\eta }%
+2Z_{0}\left( \text{A}\right) Z_{0}\left( \text{B}\right) +Z_{0}\left( \text{%
C}\right) Z_{0}\left( \text{C}\right) .  \label{Equation: St+Solv Total PF}
\end{equation}
The first term represents the contribution of the configuration with the solvent molecule at the origin, while the second one represents the contribution of the two possible configurations with the end point at the origin and the last one represents the contribution of the configuration with the core at the origin. The factor two is related to the possibility of having the star on either side of the origin when the end point is at the origin.

The solvent density is then defined as the ratio between the partition function describing configurations in which the origin is occupied by a solvent molecule and the total partition function of the system. It is easy to see that it is given by (\ref{solvent_density}).The (mass) density of the star particles is similarly given by 
\begin{equation}
\phi _{\text{st}}=\frac{2Z_{0}\left( \text{A}\right) Z_{0}\left( \text{B}%
\right) +Z_{0}\left( \text{C}\right) Z_{0}\left( \text{C}\right) }{Z_{0}}.
\end{equation}%
Since every star particle occupies five sites on the lattice, the number density of stars is 
\begin{equation}
\phi _{\text{st,n}}=\phi _{\text{st}}/5.
\end{equation}

The total partition function can be used to obtain the thermodynamics of the system with a calculation that is formally identical to the one carried on in the previous section for the case of monomeric particles and squares.  The free energy per site is still given by (\ref{FreeEnergy0}), but 
$Q_{2}$\ is the polynomial 
\begin{equation}
Q_{2}=\frac{s^{2}}{\eta }+2ab+c^{2}.
\end{equation}

As in the previous case, the calculations are done in the grand canonical ensemble and are carried out at constant chemical potentials. Hence, the above free energy represents $\beta P_{\mathrm{st}}v_{0}$ where $\beta $ is the inverse temperature, $v_{0}$ represents the volume of the unit cell of the lattice and $P_{\mathrm{st}}$ is the osmotic pressure \cite{PDG-JCP3-98} across a membrane permeable to the star particles. The osmotic pressure $P_{\mathrm{s}}$ across a membrane permeable to the solvent particles is given by $\beta P_{\mathrm{s}}v_{0}\equiv \beta P_{\mathrm{st}}v_{0}-\ln \eta.$ If the solvents represent voids, then $P_{\mathrm{s}}$ represents the conventional pressure of the lattice system \cite{PDG-JCP3-98}.

In order to determine which phase is the stable one at some temperature, we must find the free energy of all the possible phases of the system as a function of $w$.

\subsubsection{$\protect\mu >0$}

Let us first consider the case of positive chemical potential. As explained in the previous section, when $\mu $ is positive, the ground state of the lattice at zero temperature is represented by a pure solvent phase: all the lattice sites are occupied by monomeric species. As we start increasing the temperature the density of stars will increase. For positive values of $\mu $, as in the previous case, only one phase is present and, therefore, no phase transition is observed. In this case, the solvent is percolating at every temperature and what we need to investigate is the percolation of stars.

In order to study the percolation of the star particles, we introduce the probabilities $R_{m}(\alpha )\leq 1$ as above. For example, $R_{m}(C)\leq 1$ is the probability that a site in a $\text{C}$ state at the $m$th generation is connected to a finite cluster of stars at higher generations. At the 1-cycle fixed-point, each one of these $R_{m}(\alpha )$ approaches its fixed-point value given by the solution of an equation of the form  (\ref{percolation_relation}) given earlier. By considering the configurations shown in Figures \ref{Figure:star_above configurations} and \ref{Figure:star_below configurations}, and following the same steps shown
in the previous section for the square particles, we can write: 
\begin{align}
R(\text{A})& =\frac{(R(\text{A})a+w^{2}s)R(\text{B})R(\text{C})}{(a+w^{2}s)},
\\
R(\text{B})& =\frac{w^{2}s^{3}+(2w^{2}R(\text{A})+w^{4})s^{2}a+3w^{2}R(\text{%
A})^{2}sa^{2}+R(\text{A})^{3}a^{3}+R(\text{C})R(\text{B})^{2}cb^{2}}{%
w^{2}s^{3}+(2w^{2}+w^{4})s^{2}a+3w^{2}sa^{2}+a^{3}+cb^{2}}, \\
R(\text{C})& =\frac{(R(\text{A})a+w^{2}s)R(\text{B})^{2}}{(a+w^{2}s)}.
\end{align}

This system of equations does not depend explicitly on the chemical potential $\mu $ but it depends on it implicitly. In fact, for every value of $T$, and hence of $w$, a different chemical potential corresponds to different values of $s$, $a,$ $b$ and $c$ and consequently a different value for the $R(\alpha )^{\prime }$s.

Since $R(\alpha )$ represents the probability of the (occupied) site at the origin of the lattice being connected to a finite cluster of stars on one half of the lattice and since two half lattices are joined together at the origin to form the complete lattice, we define 
\begin{equation}
p_{\text{st}}=1-R(\text{C})^{2}
\end{equation}%
as the percolation probability for the stars. This quantity represents the probability to have the origin occupied by the core of a star connected to an infinite cluster of stars on both sides of the origin. As explained before \cite{firstpaper}, there are many other possible definitions for the percolation probability. The one above is the one that seems the most proper choice for this particular case. The behavior of $p_{\text{st}}$ as a function of $\phi _{\text{st}}$\ for different (positive) values of $\mu $, is shown in Figure \ref{Figure:husimi_st_solv_pos_mu_pvsphi}. As expected, the percolation threshold increases as the chemical potential increases. The athermal case corresponds to $\mu =+\infty ,\varepsilon \rightarrow 0$ such that $\mu \varepsilon $ is a positive constant.

Figure \ref{Figure:husimi_st_solv_pos_mu_phicvsmu} shows how the percolation threshold depends on the value of the chemical potential while Figure \ref{Figure:husimi_st_solv_pos_mu_tcvsmu} shows the dependence of the critical temperature. As expected, an increase in the chemical potential for the smaller particles makes it more and more difficult for the bigger particles to percolate. This explains the increase in the critical temperature and the percolation threshold as $\mu $ increases.

For any given value of the chemical potential, the percolation threshold and the critical temperature are lower in the case of star particles as compared to the case of square particles presented in the previous section. Consequently, our results describe the fact that, as observed in experimental results, an increase in the size disparity among the particles that form a physical system leads to a decrease in the percolation threshold. It is necessary to have a smaller fraction of sites occupied by the star particles, as compared to the square particles, in order to observe percolation.

\subsubsection{$\protect\mu <0$}

If negative values of $\mu $ are considered, the problem becomes more interesting and remarkably more complicated, even compared to the problem of  square and monomeric particles analyzed in the previous section. As explained in the previous section, a negative value for $\mu $ means that, according to thermodynamics, the system will be stable at low temperature if all the sites are covered by star particles.

As it happens with the system formed by monomeric and square particles, what can be expected at low temperature, or at very low concentrations of solvent  molecules, is an ordered structure of stars which disorders at high temperature, so that there is an abundance of star particles in the system. Thus, we could have a phase transition between an ordered phase, stable at low temperature and a disordered phase, stable at higher temperature. Figure \ref{Figure:husimi_st_solv_neg_mu_omegavst} shows the free energy of the system in the case of $\mu =-1$.

By using the set of equations introduced above, the free energy that one obtains is the one represented by the open circles in the figure. This cannot be the free energy corresponding to the stable phase at every temperature for the same reason as given in the previous section, since it becomes negative below some critical temperature.

In order to capture the ordered phase for this problem, it is necessary to  have a calculation scheme that goes beyond the one introduced above. This is exactly what had to be done in the case of monomeric and square particles in the previous section. The main problem here is related to the fact that even though we had a somehow clear understanding of what the geometrical nature of the ground state should have been, its mathematical description is everything but simple. If it is necessary to solve one recursive equation in one variable there are very well known theorems that provide necessary and sufficient conditions for the convergence of the recursion relation of one variable, and consequently for the solution of the problem. But, if we have a system of two or more equations in two or more variables, even as simple as possible but not linear, there are no conditions that guarantee the convergence of the recursion relations. Consequently, the solution is obtained through a process of trial and error. Unfortunately, even though a large number of different systems of equations have been solved, we have not been able to find a scheme that works for all the systems that we have considered. Even if we can immediately determine what approach will not work, for example in the case of the stars we cannot expect the one-cycle approach to be able to describe the ground state for the reasons stated above, we cannot tell a priori what recursion scheme will work. In the case of the system made of monomeric and star particles, a very complicated scheme has been used. The details of the equations used to solve the problem are not very enlightening. So the equations are not reported here. But here 
we will give the rationale behind the choice of this particular set of recursion relations for this problem.

The key is to distinguish the directions along which one moves on the lattice and the position along which a star is placed inside a given plaquette. Figure \ref{Figure:husimi_st_1234labeling} shows how this can be achieved. We distinguish the directions labeled 1, 2, 3 and 4 in the figure because of the very peculiar nature of the ground state. It is necessary to use this description in order to be able to capture the ground state that has a very regular pattern and where, as explained above in the case of the percolation of square particles, at low temperature when the lattice is almost completely covered by stars there are very strong correlations between sites that are second or third neighbors on the lattice. In this case, in order to get a solution it was necessary to introduce a symmetry breaking field to make the ground state unique. The symmetry breaking field makes some direction more favored than the others, thus creating a unique ground state for the system. There are four different fixed points that come out from the recursion relations. These different fixed points have to be carefully accounted for in order to obtain the proper free energy for this system.

In this case, the approach is more complicated than in the case of the square particles. It is necessary to use a \textit{multi-cycle method}. In this case, what is checked is the difference of the values of all the ratios at two levels that are \textit{n}-generations apart on the lattice. Whenever $\alpha _{m}-\alpha _{m+n}$ for all the ratios $\alpha $ is less than the desired tolerance we assume the fixed point of the solutions has been reached. Wherever the ordered phase exists (at low temperature) for the case of stars on a square Husimi lattice, we have a \textit{four-cycle} structure of the fix-point, so that the quantity $\alpha _{m}-\alpha _{m+4}$ is the one that is going to vanish.

By using this second method, it is possible to obtain a second solution for the problem. The corresponding free energy curve is the one represented by the circles in Figure \ref{Figure:husimi_st_solv_neg_mu_omegavst}. The solutions obtained with the two different schemes coincide at high temperatures, but are different at low temperatures. The temperature at which the ordered phase appears and becomes the stable solution is, in the case of $\mu =-1$, $T_{\mathrm{OD}} \simeq 8.4$. We use the subscript OD, as done before in the case of the square particles, to remind that this transition is a phase transition between an ordered and a disordered phase. It is possible to observe that the stable phase at every temperature has a positive free energy, as expected.

The entropy corresponding to the free energy shown in Figure \ref{Figure:husimi_st_solv_neg_mu_omegavst} is shown in Figure \ref{Figure:husimi_st_solv_neg_mu_entropyvst}. As expected, below $T_{\text{OD}}$ the entropy of the ordered phase is lower than the entropy of the metastable continuation of the disordered phase. The entropy of the metastable phase, though, drops much more rapidly, and goes to zero at a finite temperature. This temperature is the Kauzmann temperature of the system. 

The results for the percolation probability as a function of the temperature are shown in Figure \ref{Figure:husimi_st_solv_neg_mu_pvst}. It is obvious that the strength of the percolation process is very different for the two phases. The formation of a percolating cluster of solvent molecules is much easier in the disordered phase than in the ordered one. There is a wide temperature range, between $\sim 2.70$ and $\sim 7.45$ in the case $\mu =-1$, in which there is at least one percolating cluster in the disordered phase but no percolation occurs in the ordered phase. The region below $2.70$, limited by the dashed line in Figure \ref{Figure:husimi_st_solv_neg_mu_pvst} does not represent physical states of the disordered phase because it is below the Kauzmann temperature and it corresponds to a negative entropy. This figure shows how the percolation threshold for the two phases is very different. In particular the solvent density must be at least $\sim 0.32$ in order to have percolation in the ordered phase while it is much lower in the metastable extension of the disordered phase. During the lowering of the temperature, if the phase transition is avoided, the system will percolate all the way down to the Kauzmann temperature. Figure \ref{Figure:husimi_st_solv_neg_mu_tcvsmu} summarizes the results obtained as the value of the solvent chemical potential is changed.

\section{Conclusions}

If we compare the results described above, we can draw some interesting conclusions about the effect of the size and shape of the particles on the percolation properties of these systems.

The two systems that have been considered on the Husimi lattice have one common species, the monomeric one, and a different one, square particles in the first case and star particles in the second case.

Both systems present common features. For instance, a metastable phase is present that is percolating very strongly in a temperature interval where the stable phase has a very weak percolating cluster (a much lower value of $p$) or even no percolating cluster at all. This observation suggests how it might be useful to prepare a system in a metastable state in some cases in which a percolating network of particles of some kind is needed in order to enhance one or more physical properties of the composite system. The presence of the star particles lowers the percolation threshold that decreases as the size disparity between different particles present in the system increases since the each star particle occupies more sites than a square particle. The introduction of the star particles, as compared to the square particles has an effect on the thermodynamics of the system as well. If we consider the case of negative values of the monomeric species chemical potential, we observe in both cases a continuous transition between an ordered phase that is stable at high temperature and a disordered phase that is instead stable above a critical temperature. The system that contains the star particles, though, has a much lower value of the critical temperature, as compared to the system containing the square particles. The presence of the larger star particles makes the disordering of the system easier as compared to the case in which the smaller square particles are present. So the system has a critical temperature that is lower in the case of the presence of star particles than in the case of the square particles. 

We have recently also looked at the percolation of stars and solvent particles on a Bethe lattice \cite{Corsi-PDG-Bethe-tbp}. In this case, the most striking difference, as compared to the case of the Husimi lattice, is the different nature of the percolation transition since on the Bethe lattice we obtain a discontinuous transition for the percolation probability. At this stage of the research, the results obtained do not allow us to draw any conclusions about the reason of the different nature of the phase transition between the ordered and the disordered phase on the two different lattice structures that we have used, the Husimi lattice and the Bethe lattice. We believe that the size disparity coupled with the nature of the lattice plays an important role in determining the nature of the transition.  Further investigations are needed to give a complete answer to this question and they will be pursued in the future. In particular, it would be very interesting to study the behavior of larger and more complicated particles on both a Husimi lattice and a Bethe lattice. Of course, an increase in the complexity of the particles that are treated would likely lead to a tremendous increase in the complexity of the recursive equations that need to be solved. This would represent a problem mainly for the description of the ground state of the phase rich in larger particles. 

Now that we are able to describe the properties of systems made of particles of different sizes and shapes, we move forward and include the effect of having a polymer matrix in which the particles are embedded \cite{thirdpaper}.

\newpage

\bibliographystyle{unsrt}
\bibliography{bio}

\newpage

\begin{figure}
FIGURE CAPTIONS

\caption{Portion of an infinite lattice known as Husimi tree.\hfill }
\label{Husimi}

\caption{Possible states for the $m$th level site in the case of percolation of solvent and square particles. A, B and S refer to the three possible states of the base site of the plaquette, see text for details.}
\label{Figure:square+solvent}

\caption{Percolation of squares and solvent molecules: possible configurations of the sites in the $m$th level square when the base is in the $S$ state.}
\label{Figure:square_solvent configurations}

\caption{Percolation of squares and solvent molecules: possible configurations of the sites in the $m$th level square when the base is in the $B$ state.}
\label{Figure:square_below configurations}

\caption{Percolation of squares and solvent molecules: possible configurations of the sites in the $m$th level square when the base is in the $A$ state.}
\label{Figure:square_above configurations}

\caption{Probability that a square particle at the origin of the lattice is connected to an infinite cluster of square particles spanning the entire lattice as a function of the density of squares.}
\label{Figure:husimi_sq_solv_pos_mu_pvsphi}

\caption{Probability that a square particle at the origin of the lattice is connected to an infinite cluster of square particles spanning the entire lattice as a function of the temperature of the system.}
\label{Figure:husimi_sq_solv_pos_mu_pvst}

\caption{Percolation threshold as a function of the chemical potential for the percolation of squares on a Husimi tree filled with square and solvent particles.}
\label{Figure:husimi_sq_solv_pos_mu_phicvsmu}

\caption{Critical temperature as a function of the chemical potential for the percolation of squares on a Husimi tree filled with square and solvent particles.}
\label{Figure:husimi_sq_solv_pos_mu_tcvsmu}

\caption{Free energy as a function of temperature for $\protect\mu =-1$. See text for details. \hfill}
\label{Figure:husimi_sq_solv_neg_mu_omegavst}

\caption{Entropy as a function of temperature for $\protect\mu =-1$. See text for details. \hfill}
\label{Figure:husimi_sq_solv_neg_mu_entropyvst}

\end{figure}

\newpage

\begin{figure}

\caption{Solvent percolation probability as a function of temperature for $\protect \mu =-1$ in the presence of square particles. The dotted line corresponds to the Kauzmann temperature.}
\label{Figure:husimi_sq_solv_neg_mu_pvst}

\caption{Solvent percolation probability as a function of solvent density for $
\protect\mu=-1$ in the presence of square particles. }
\label{Figure:husimi_sq_solv_neg_mu_pvsphis}

\caption{Critical temperature as a function of the chemical potential for the percolation of monomeric filler particles on a Husimi tree filled with square and solvent particles.}
\label{Figure:husimi_sq_solv_neg_mu_tcvsmu}

\caption{Percolation threshold as a function of the chemical potential for the percolation of monomeric filler particles on a Husimi tree filled with square and solvent particles.}
\label{Figure:husimi_sq_solv_neg_mu_phicvsmu}

\caption{Possible states for the $m$th level site in the case of percolation of solvent and star particles. A, B, C and S refer to the four possible states of the base site of the plaquette, see text for details..}
\label{Figure:star+solvent}

\caption{Percolation of stars and solvent molecules: possible configurations of the sites in the $m$th level square when the base is in the $S$ state.}
\label{Figure:star_solvent configurations}

\caption{Percolation of stars and solvent molecules: possible configurations
of the sites in the $m$th level square when the base is in the $A$ state.}
\label{Figure:star_above configurations}

\caption{Percolation of stars and solvent molecules: possible configurations
of the sites in the $m$th level square when the base is in the $B$ state.}
\label{Figure:star_below configurations}

\caption{Percolation of stars and solvent molecules: possible configurations
of the sites in the $m$th level square when the base is in the $C$ state.}
\label{Figure:star_core configurations}

\caption{Probability that a star particle at the origin of the lattice is connected to an infinite cluster of star particles spanning the entire lattice as a function of the density of stars.}
\label{Figure:husimi_st_solv_pos_mu_pvsphi}

\end{figure}

\begin{figure}

\caption{Probability that a star particle at the origin of the lattice is connected to an infinite cluster of star particles spanning the entire lattice as a function of the temperature of the system.}
\label{Figure:husimi_st_solv_pos_mu_pvst}

\caption{Percolation threshold for star particles as a function of the chemical potential for the percolation of stars on a Husimi tree filled with star and solvent particles.}
\label{Figure:husimi_st_solv_pos_mu_phicvsmu}

\caption{Critical temperature as a function of the chemical potential for the percolation of stars on a Husimi tree filled with star and solvent particles.}
\label{Figure:husimi_st_solv_pos_mu_tcvsmu}

\caption{Monomeric and star particles: free energy as a function of temperature for $\protect\mu =-1$. \hfill}
\label{Figure:husimi_st_solv_neg_mu_omegavst}

\caption{Labeling of the tree used to obtain the ground state of the system for the case of star and solvent particles on a Husimi tree with a ground state made of a pure star phase.}
\label{Figure:husimi_st_1234labeling}

\caption{Monomeric and star particles: entropy as a function of temperature for $\protect\mu =-1$. \hfill}
\label{Figure:husimi_st_solv_neg_mu_entropyvst}

\caption{Monomeric and star particles on a Husimi tree: solvent percolation probability as a function of temperature for $\protect\mu =-1$. The dashed line corresponds to the Kauzmann temperature.}
\label{Figure:husimi_st_solv_neg_mu_pvst}

\caption{Monomeric and star particles on a Husimi tree: critical temperature as a function of the
chemical potential for the percolation of monomeric filler particles on a
Husimi tree.}
\label{Figure:husimi_st_solv_neg_mu_tcvsmu}

\end{figure}

\newpage

\begin{figure}[tbp]
\begin{center}
\epsfig{file=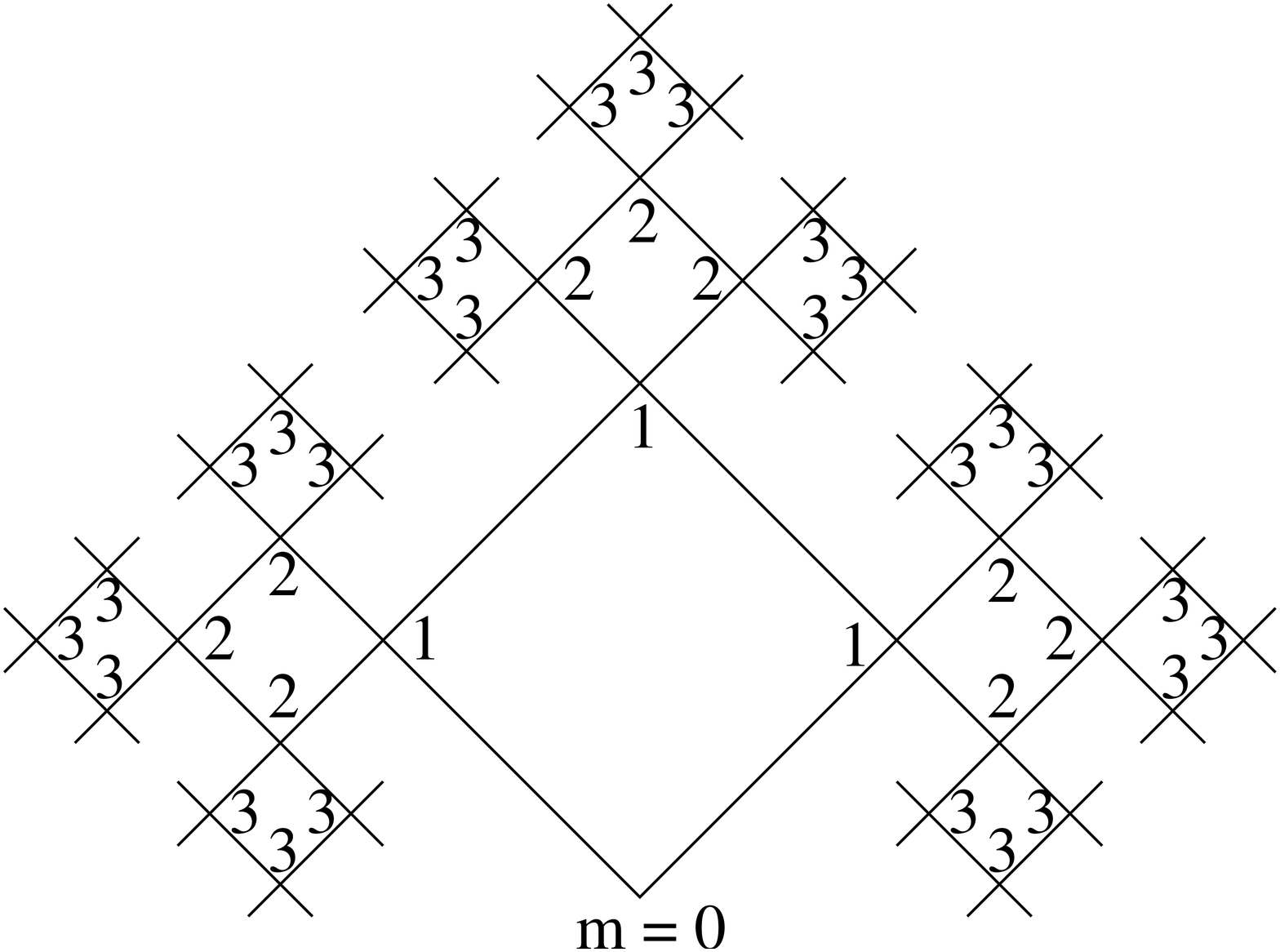,width=4in}
\end{center}
\par
FIG. 1 \vspace{5cm}
\end{figure}

\newpage

\begin{figure}[tbp]
\begin{center}
\epsfig{file=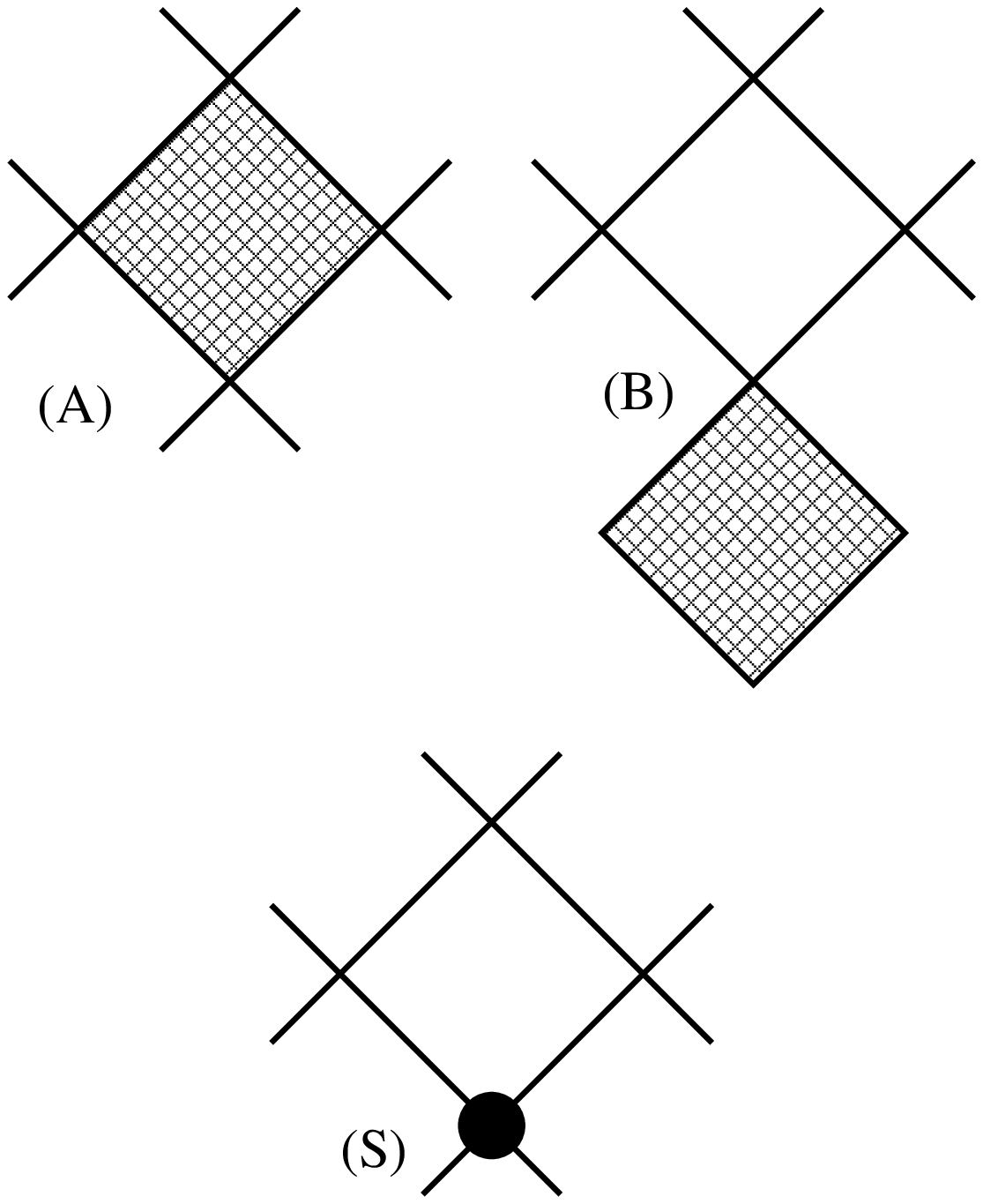,width=4in}
\end{center}
\par
FIG. 2 \vspace{5cm}
\end{figure}

\newpage

\begin{figure}[tbp]
\begin{center}
\epsfig{file=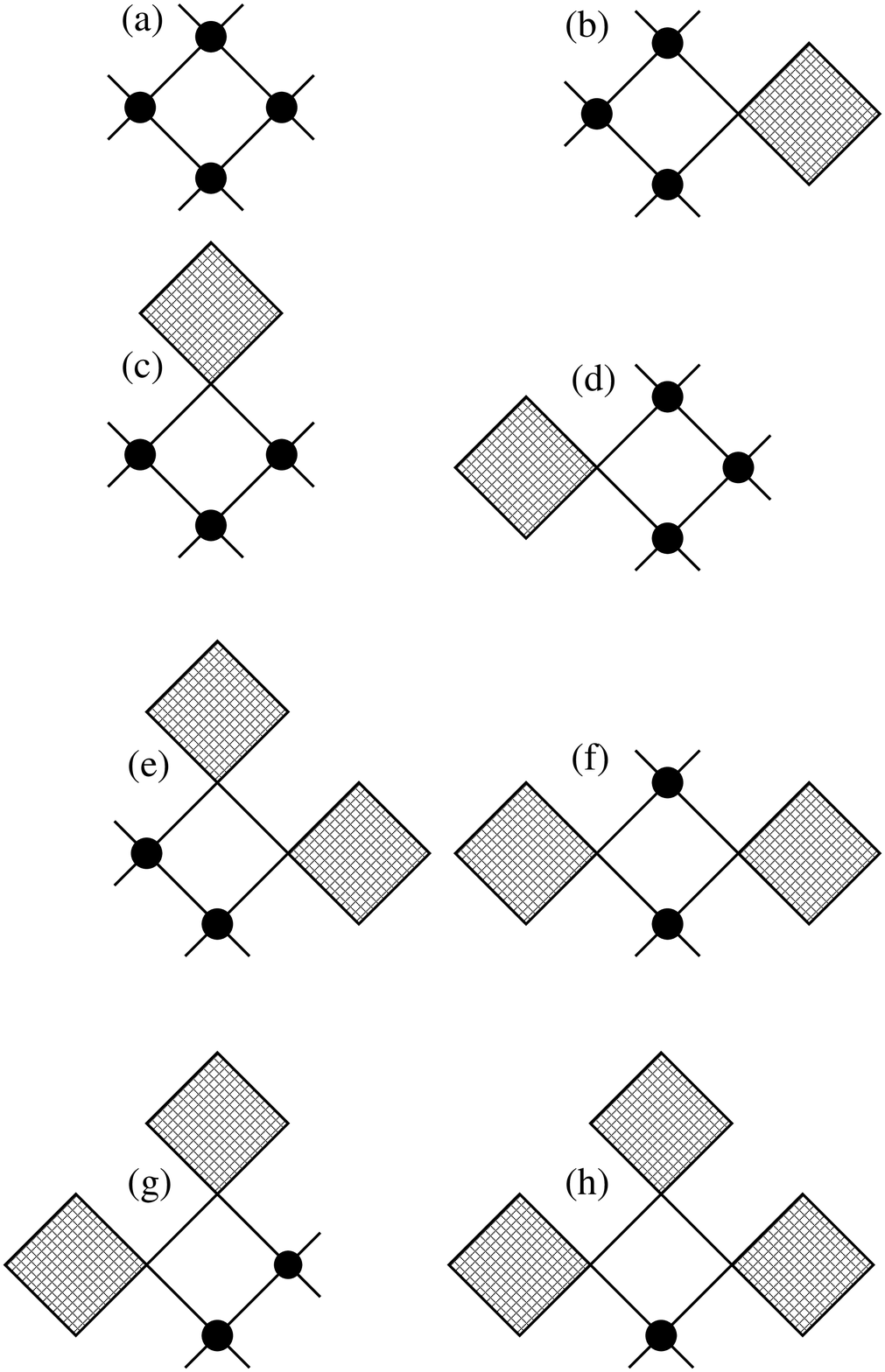,width=3.3in}
\end{center}
\par
FIG. 3 \vspace{5cm}
\end{figure}

\newpage

\begin{figure}[tbp]
\begin{center}
\epsfig{file=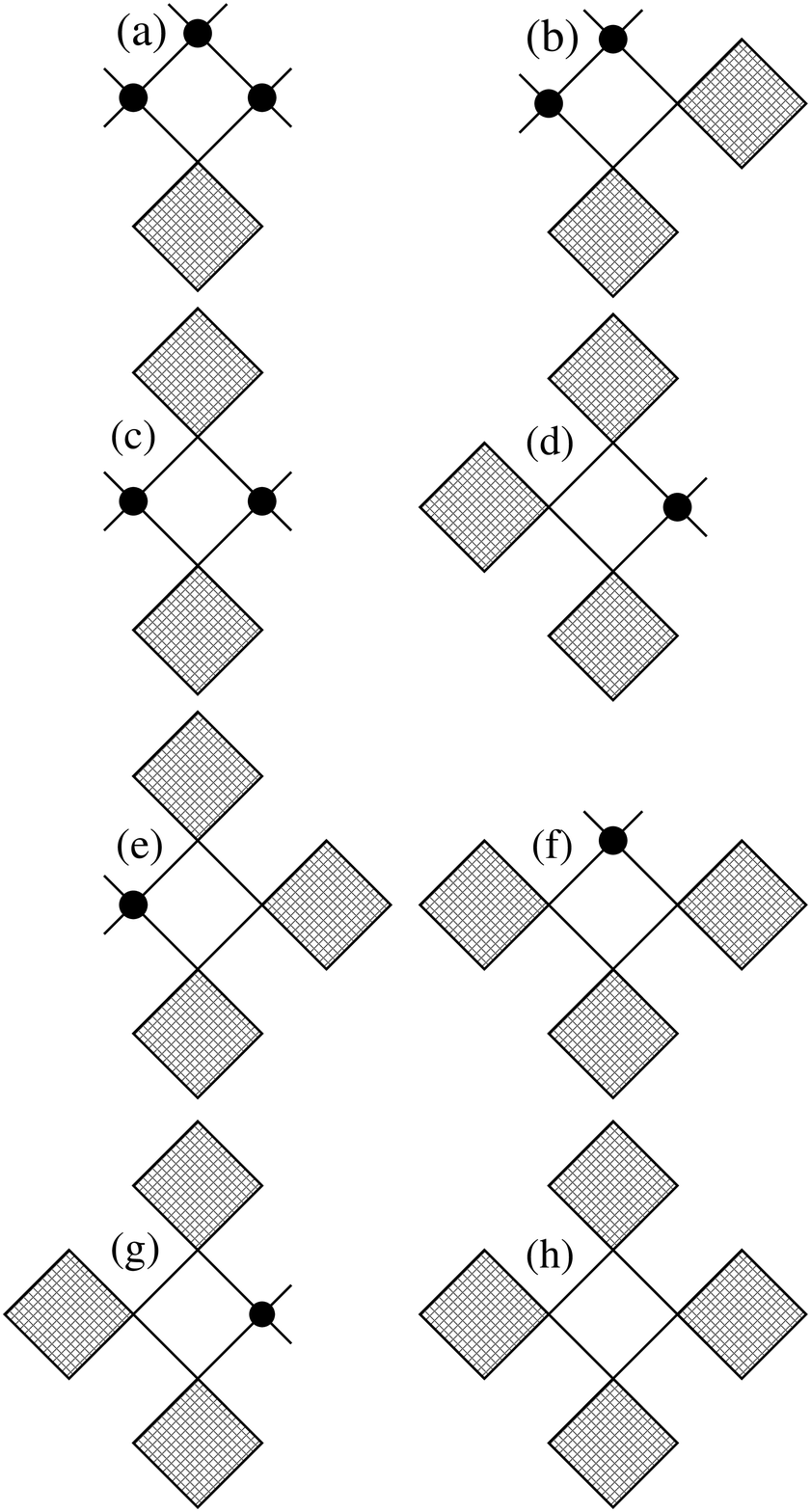,width=3.3in}
\end{center}
\par
FIG. 4 \vspace{5cm}
\end{figure}

\newpage

\begin{figure}[tbp]
\begin{center}
\epsfig{file=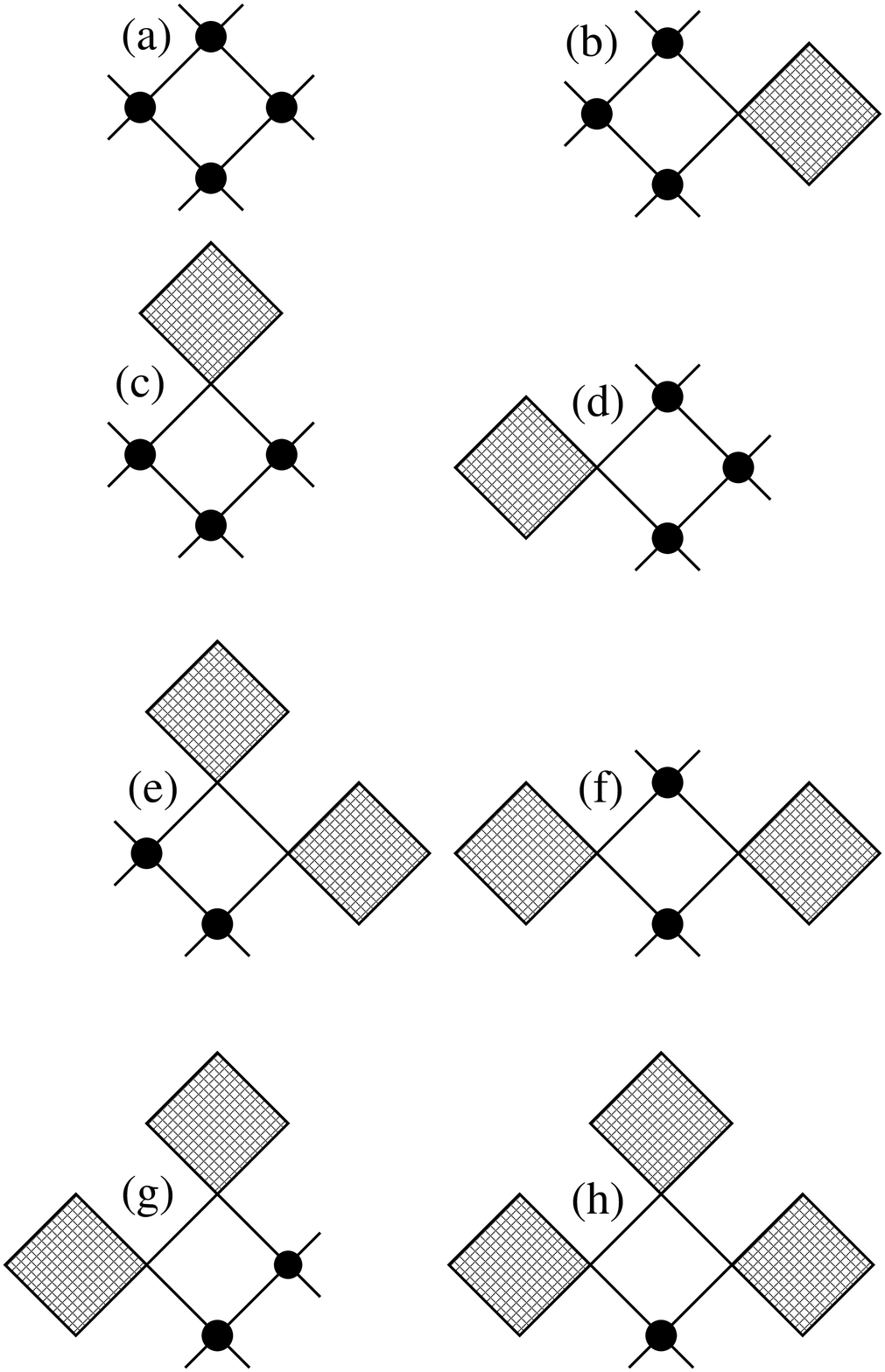,width=2in}
\end{center}
\par
FIG. 5 \vspace{5cm}
\end{figure}

\newpage

\begin{figure}[tbp]
\begin{center}
\epsfig{file=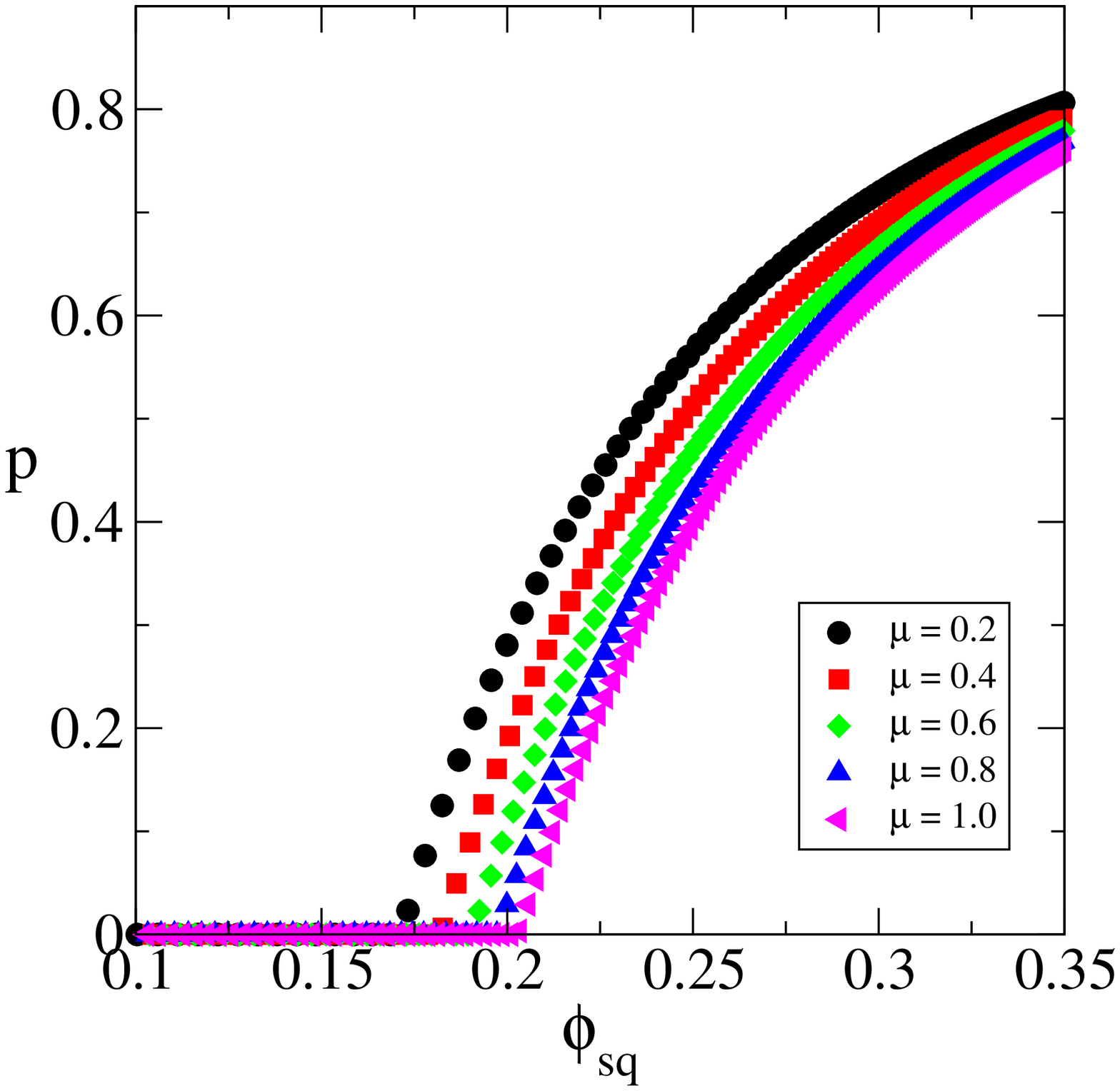,width=8.5cm,angle=270}
\end{center}
\par
FIG. 6 \vspace{5cm}
\end{figure}

\newpage

\begin{figure}[tbp]
\begin{center}
\epsfig{file=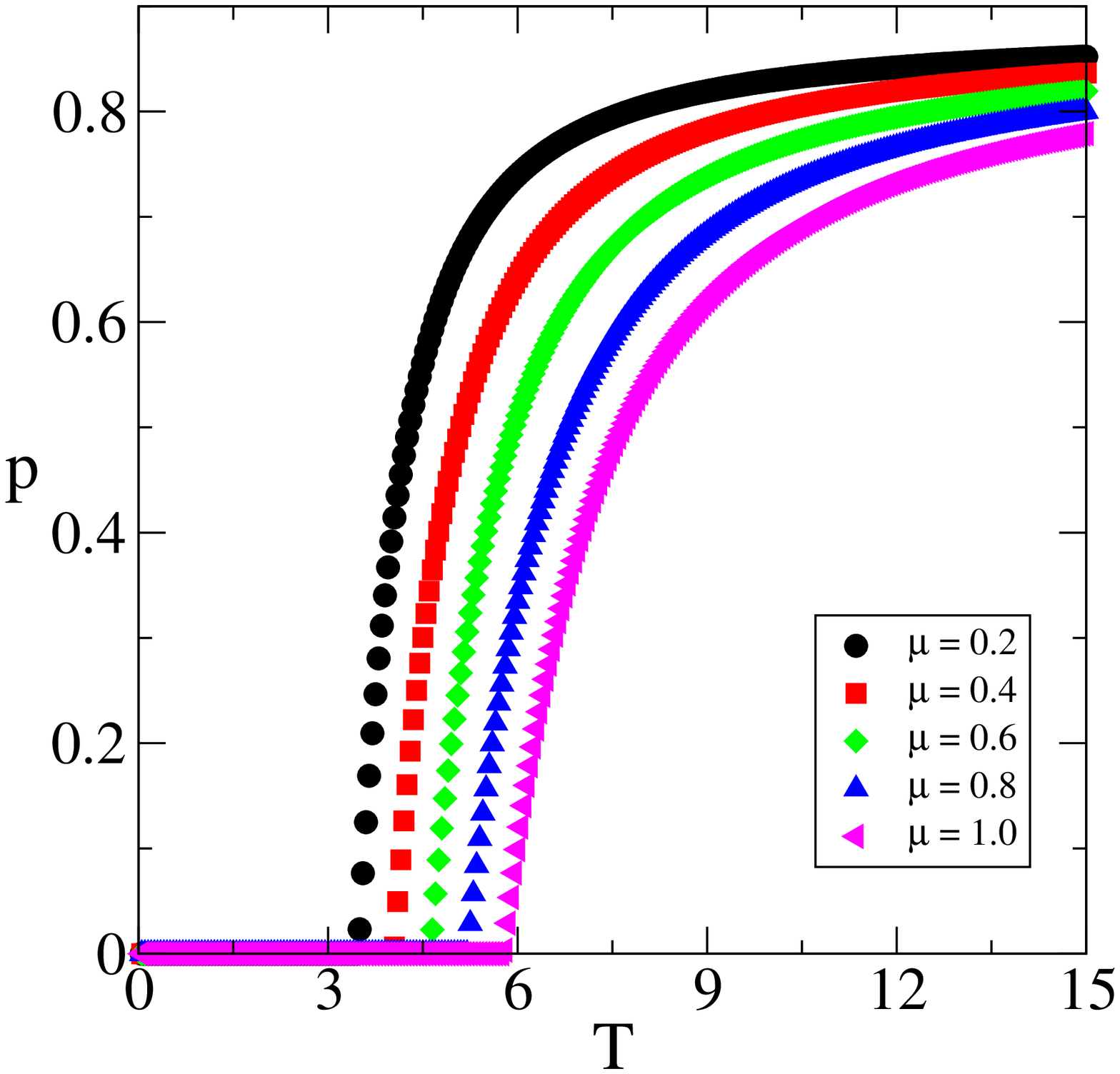,width=8.5cm,angle=270}
\end{center}
\par
FIG. 7 \vspace{5cm}
\end{figure}

\clearpage

\newpage

\begin{figure}[tbp]
\begin{center}
\epsfig{file=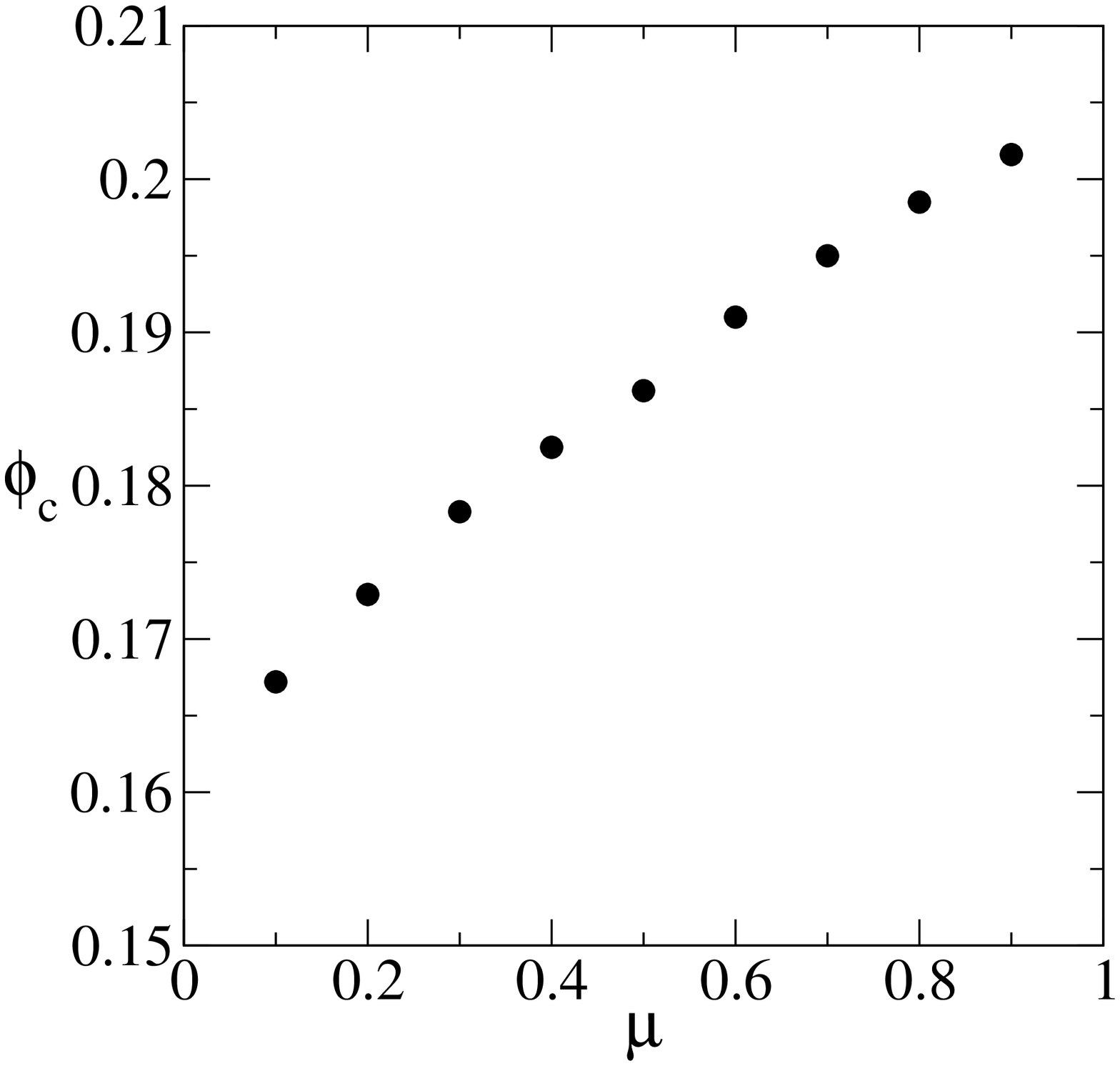,width=8.5cm,angle=270}
\end{center}
\par
FIG. 8 \vspace{5cm}
\end{figure}

\newpage

\begin{figure}[tbp]
\begin{center}
\epsfig{file=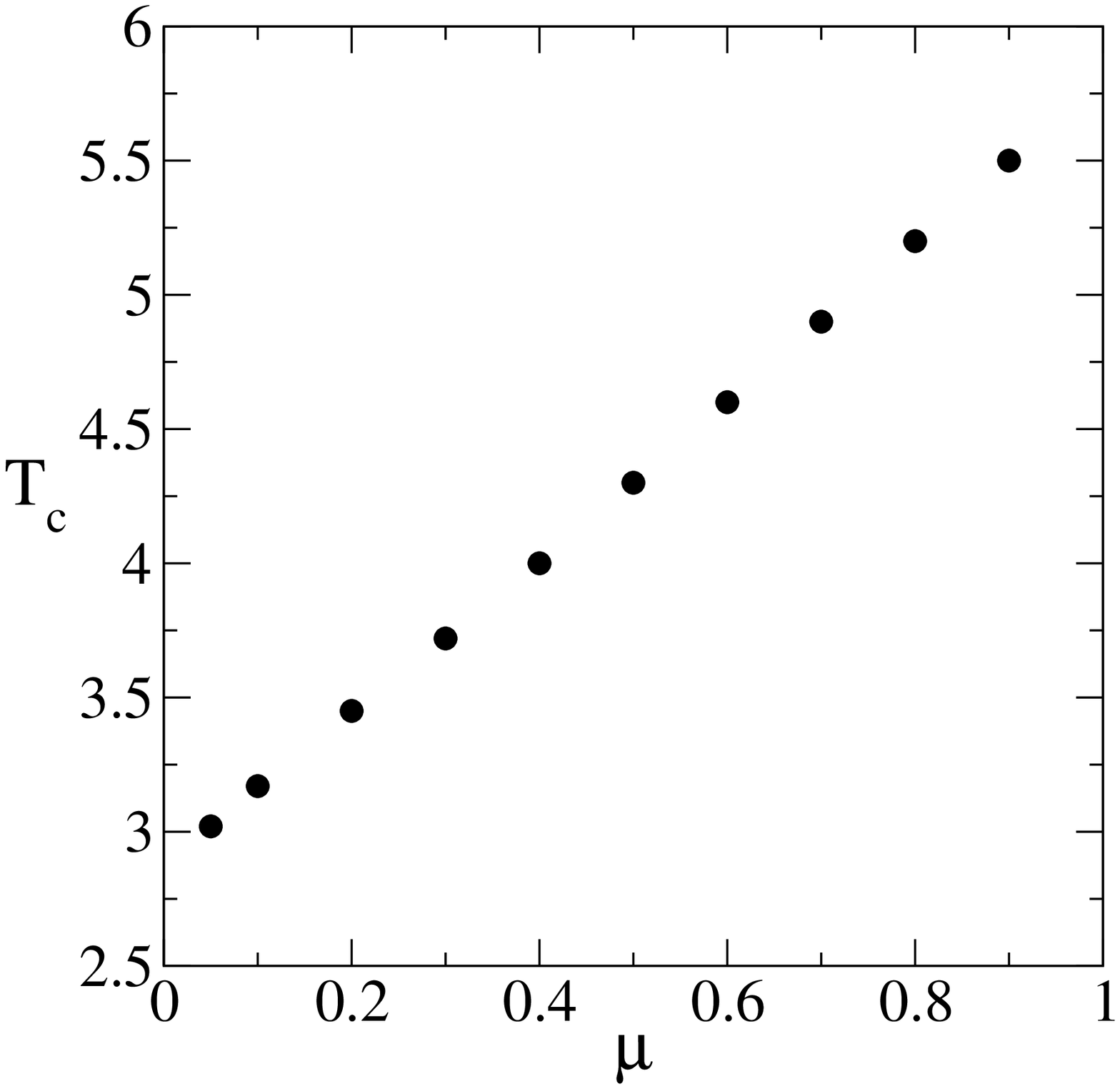,width=8.5cm,angle=270}
\end{center}
\par
FIG. 9 \vspace{5cm}
\end{figure}

\newpage

\begin{figure}[p]
\begin{center}
\epsfig{file=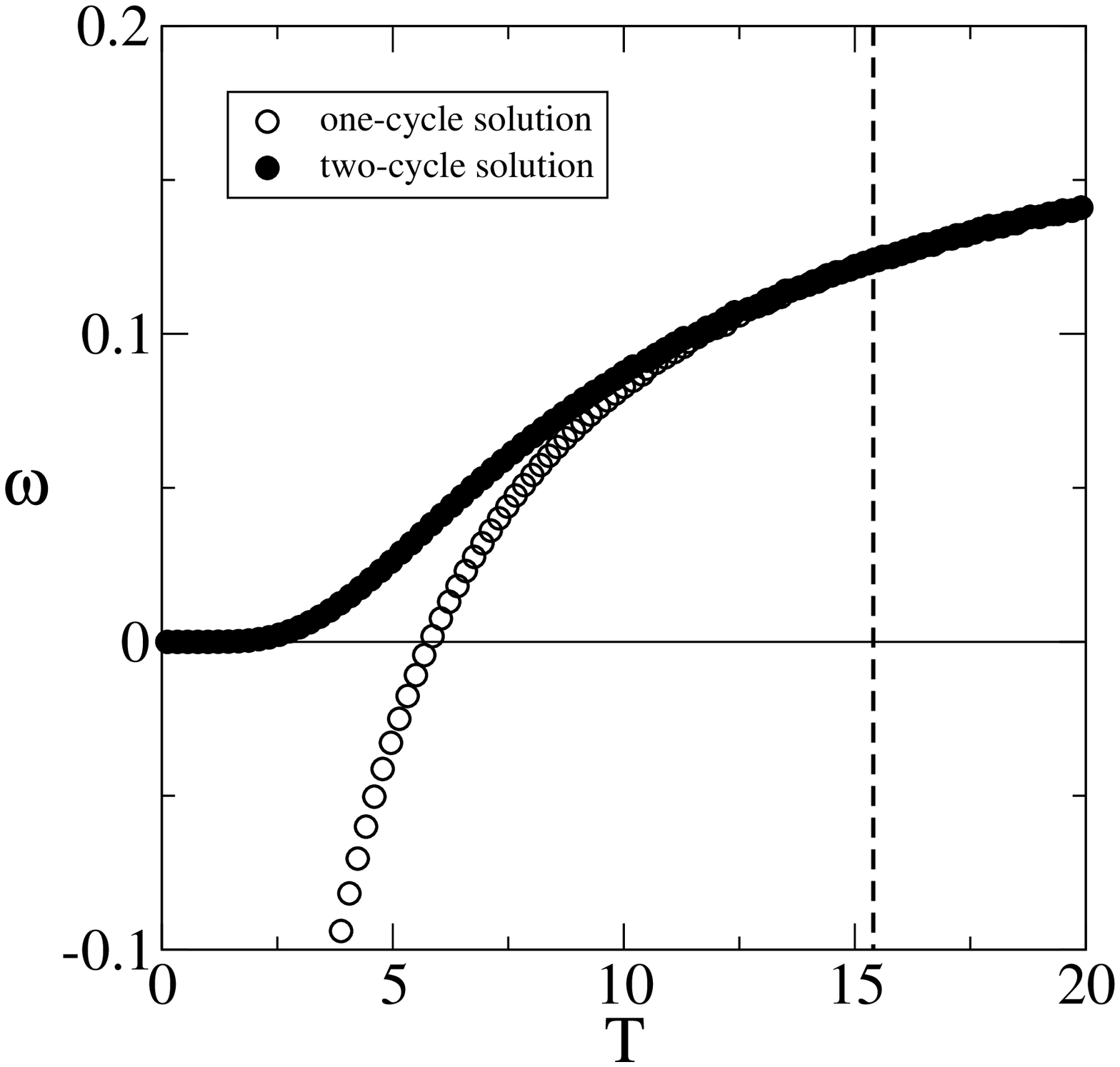,width=8.5cm,angle=270}
\end{center}
\par
FIG. 10 \vspace{5cm}
\end{figure}

\newpage

\begin{figure}[p]
\begin{center}
\epsfig{file=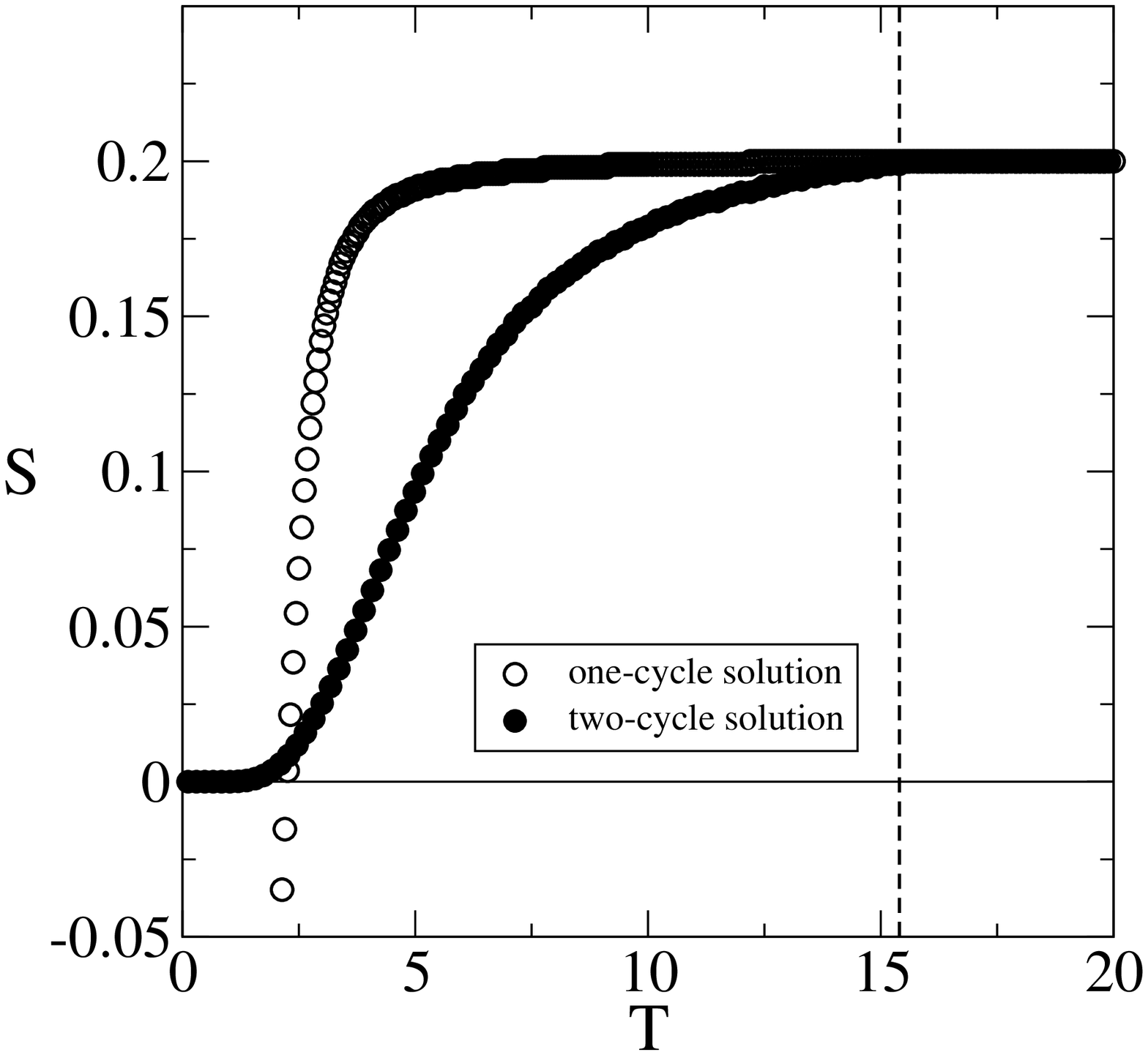,width=8.5cm,angle=270}
\end{center}
\par
FIG. 11 \vspace{5cm}
\end{figure}

\newpage

\begin{figure}[p]
\begin{center}
\epsfig{file=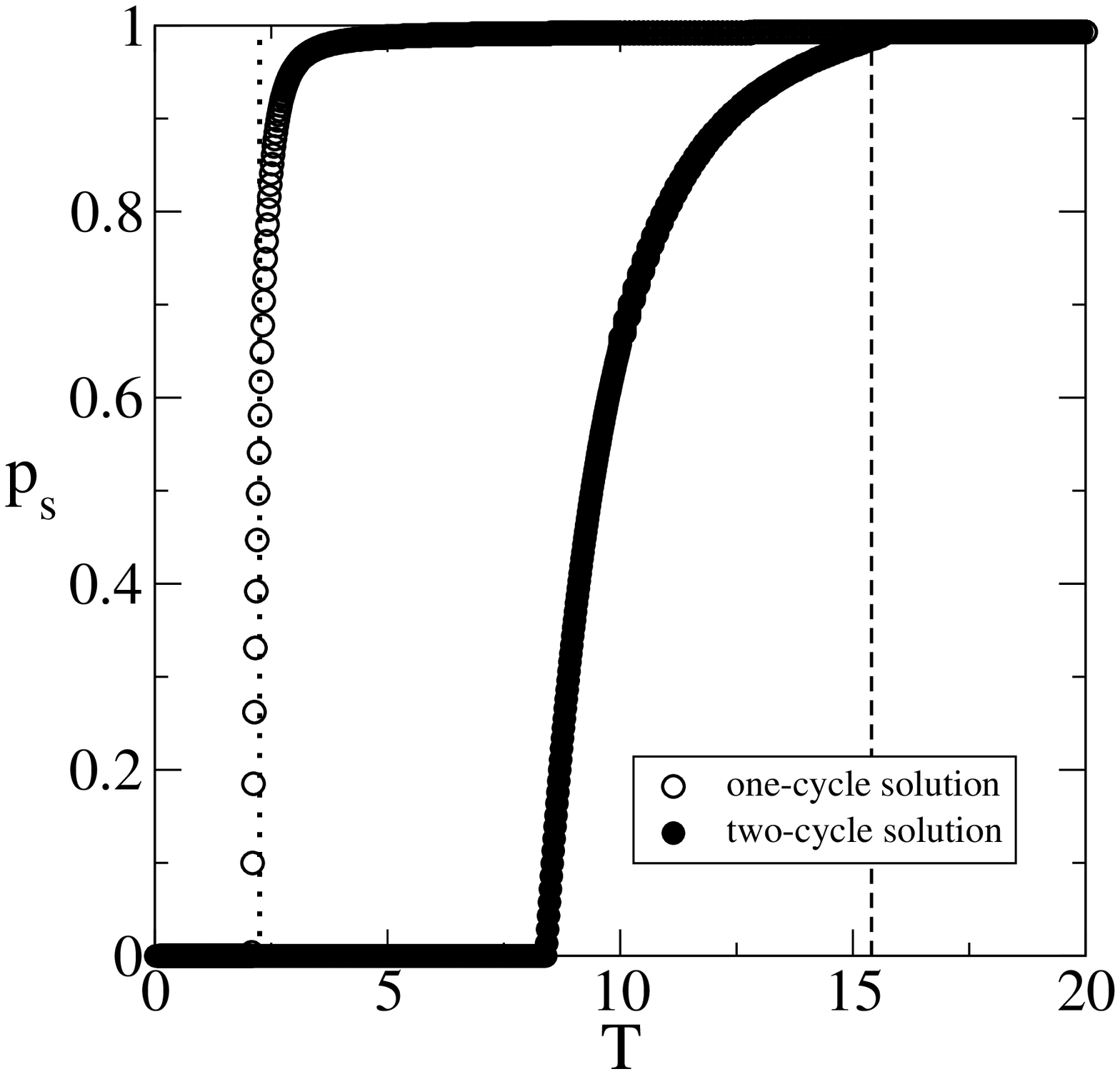,width=8.5cm,angle=270}
\end{center}
\par
FIG. 12 \vspace{5cm}
\end{figure}

\newpage

\begin{figure}[p]
\begin{center}
\epsfig{file=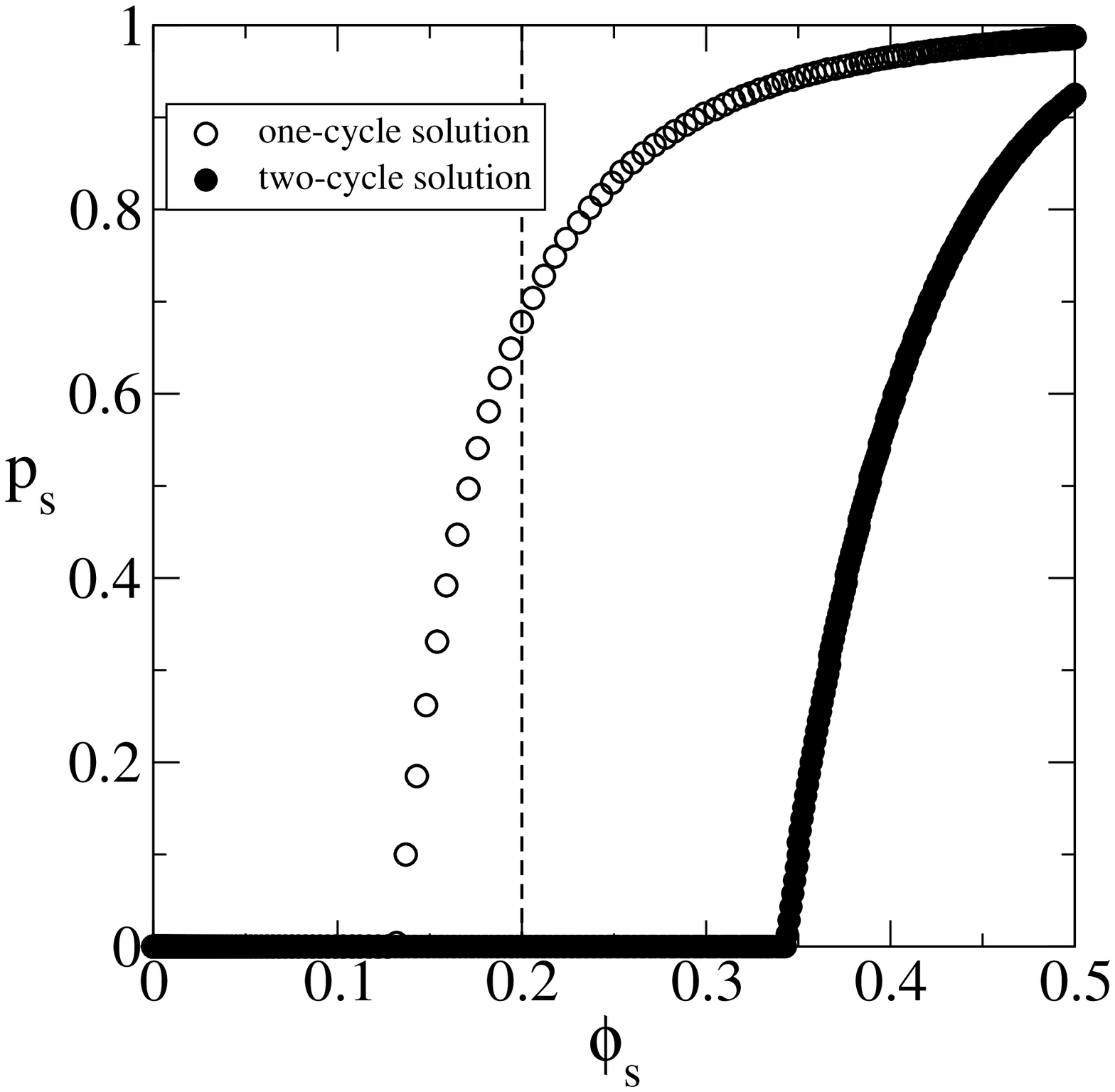,width=8.5cm,angle=270}
\end{center}
\par
FIG. 13 \vspace{5cm}
\end{figure}

\newpage

\begin{figure}[p]
\begin{center}
\epsfig{file=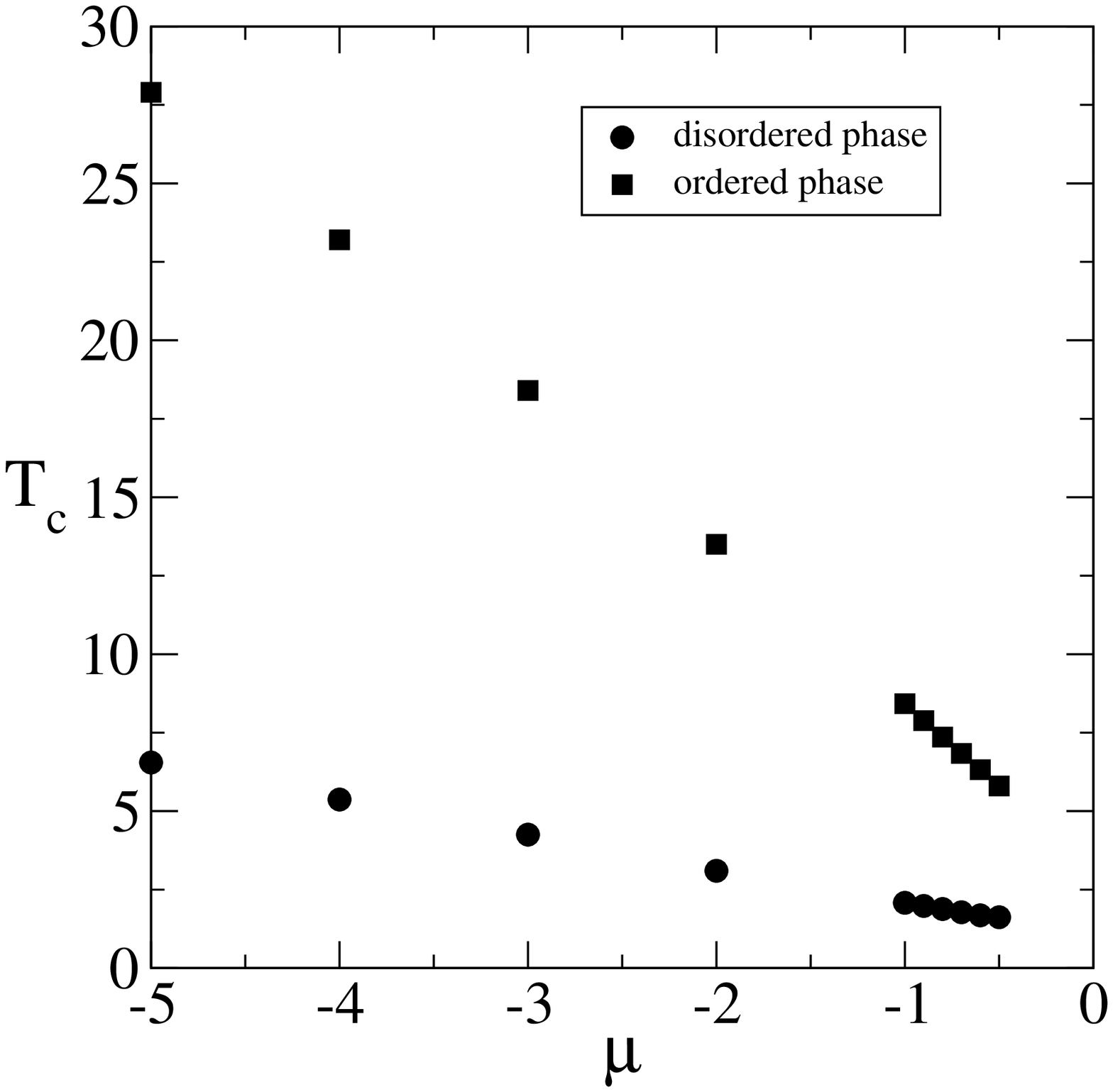,width=8.5cm,angle=270}
\end{center}
\par
FIG. 14 \vspace{5cm}
\end{figure}

\newpage

\begin{figure}[p]
\begin{center}
\epsfig{file=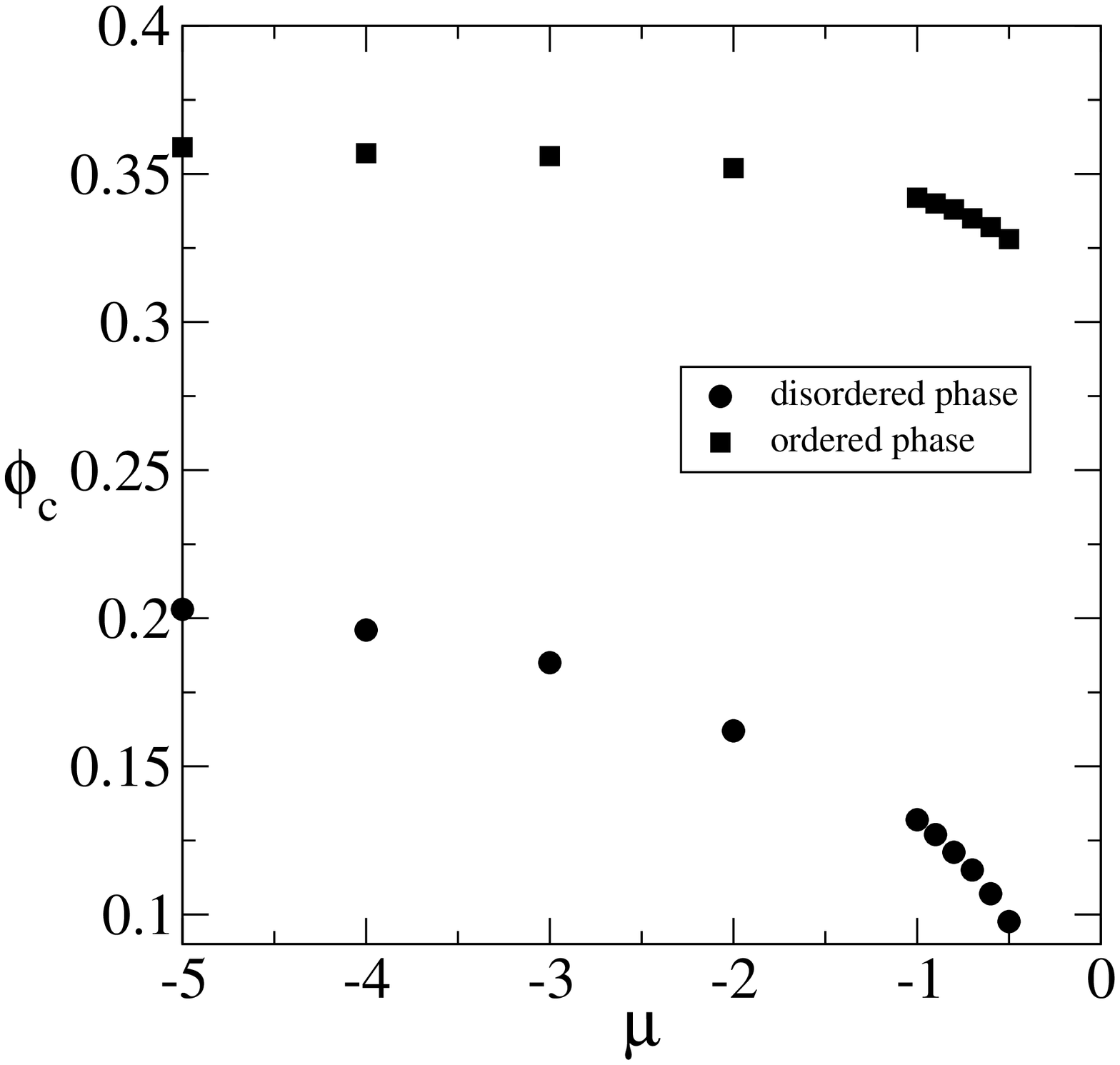,width=8.5cm,angle=270}
\end{center}
\par
FIG. 15 \vspace{5cm}
\end{figure}

\newpage

\begin{figure}[tbp]
\begin{center}
\epsfig{file=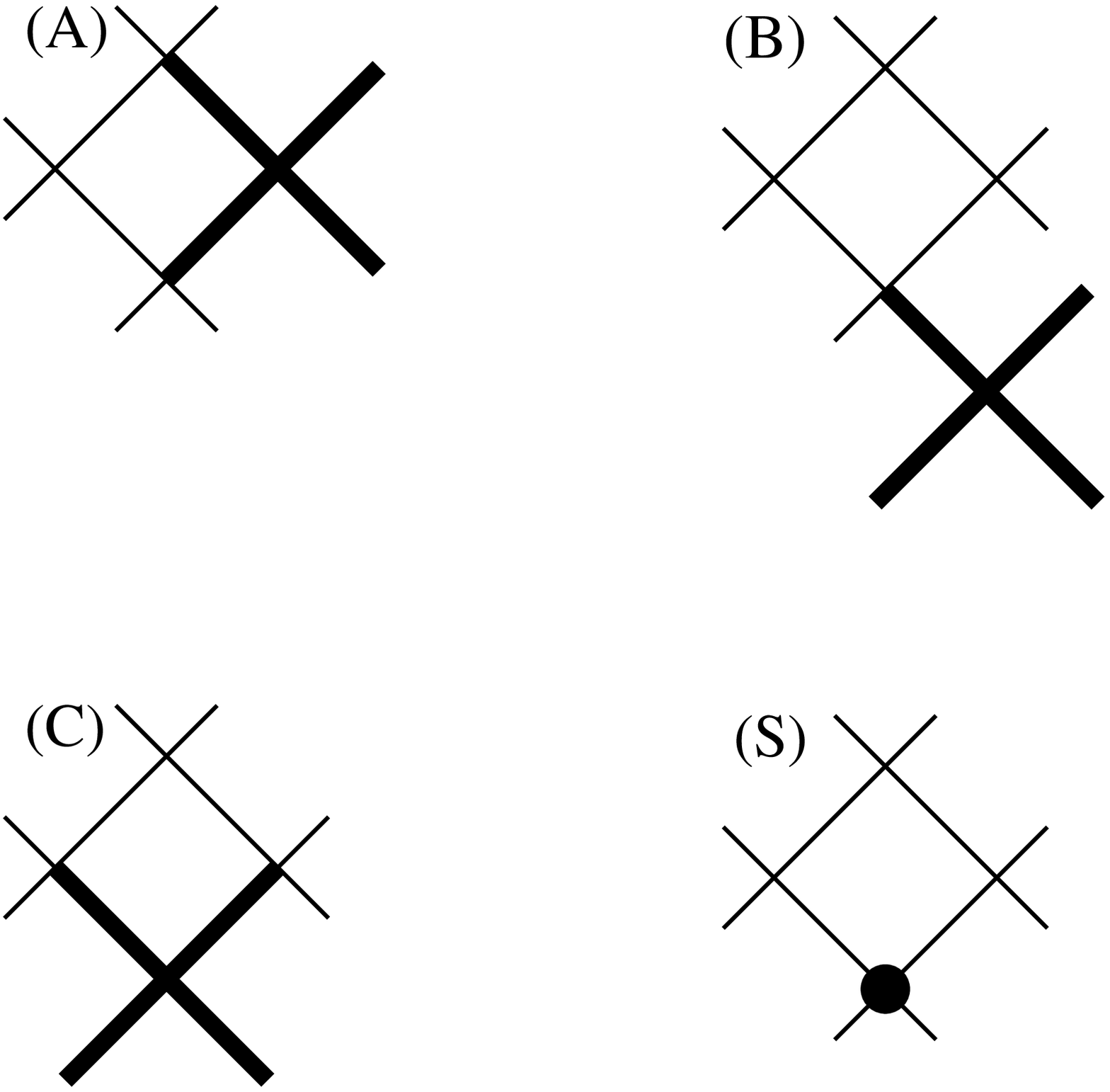,width=3in}
\end{center}
\par
FIG. 16 \vspace{5cm}
\end{figure}
\newpage

\clearpage

\begin{figure}[tbp]
\begin{center}
\epsfig{file=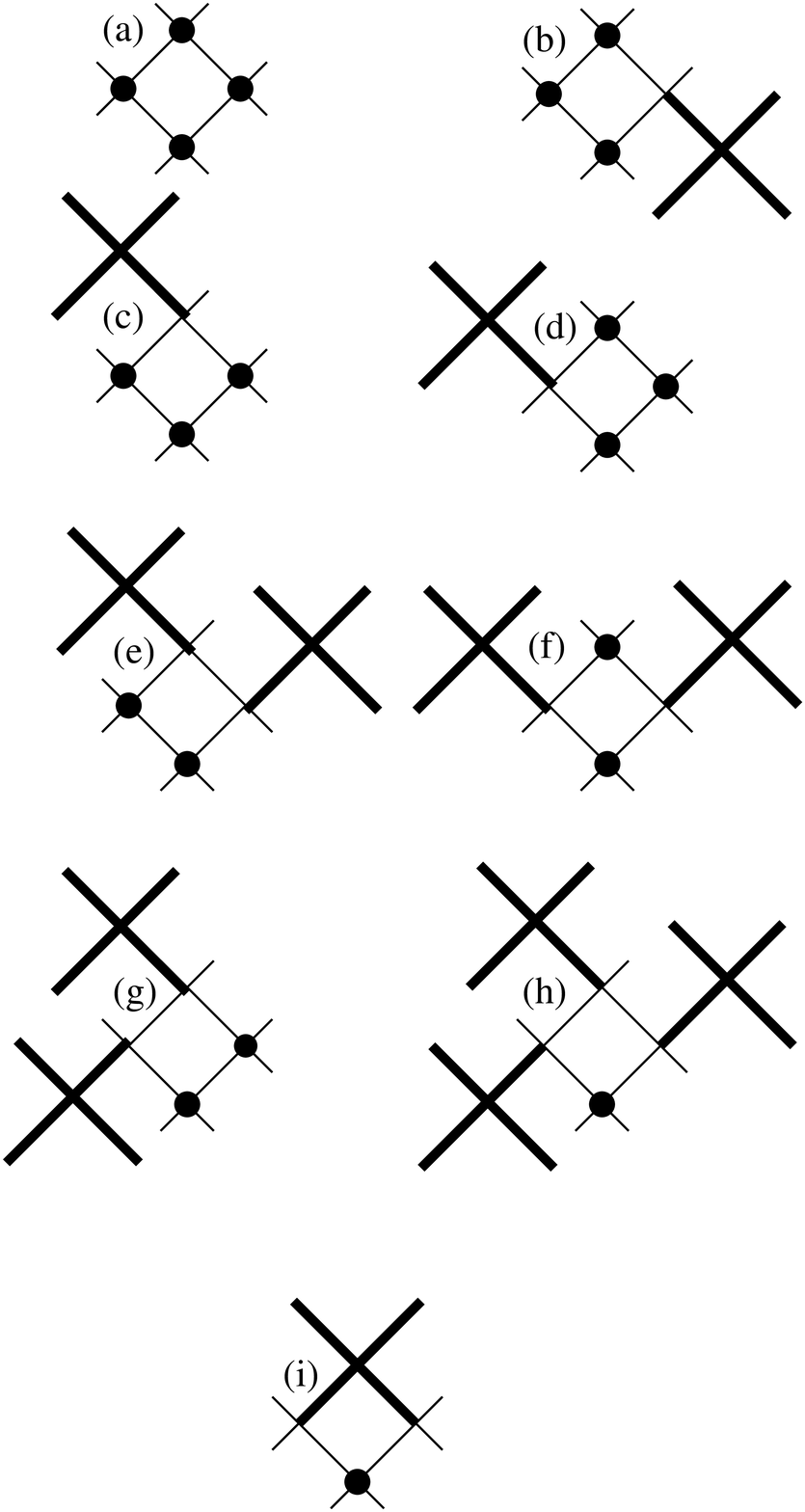,width=3in}
\end{center}
\par
FIG. 17 \vspace{5cm}
\end{figure}

\newpage

\begin{figure}[p]
\begin{center}
\epsfig{file=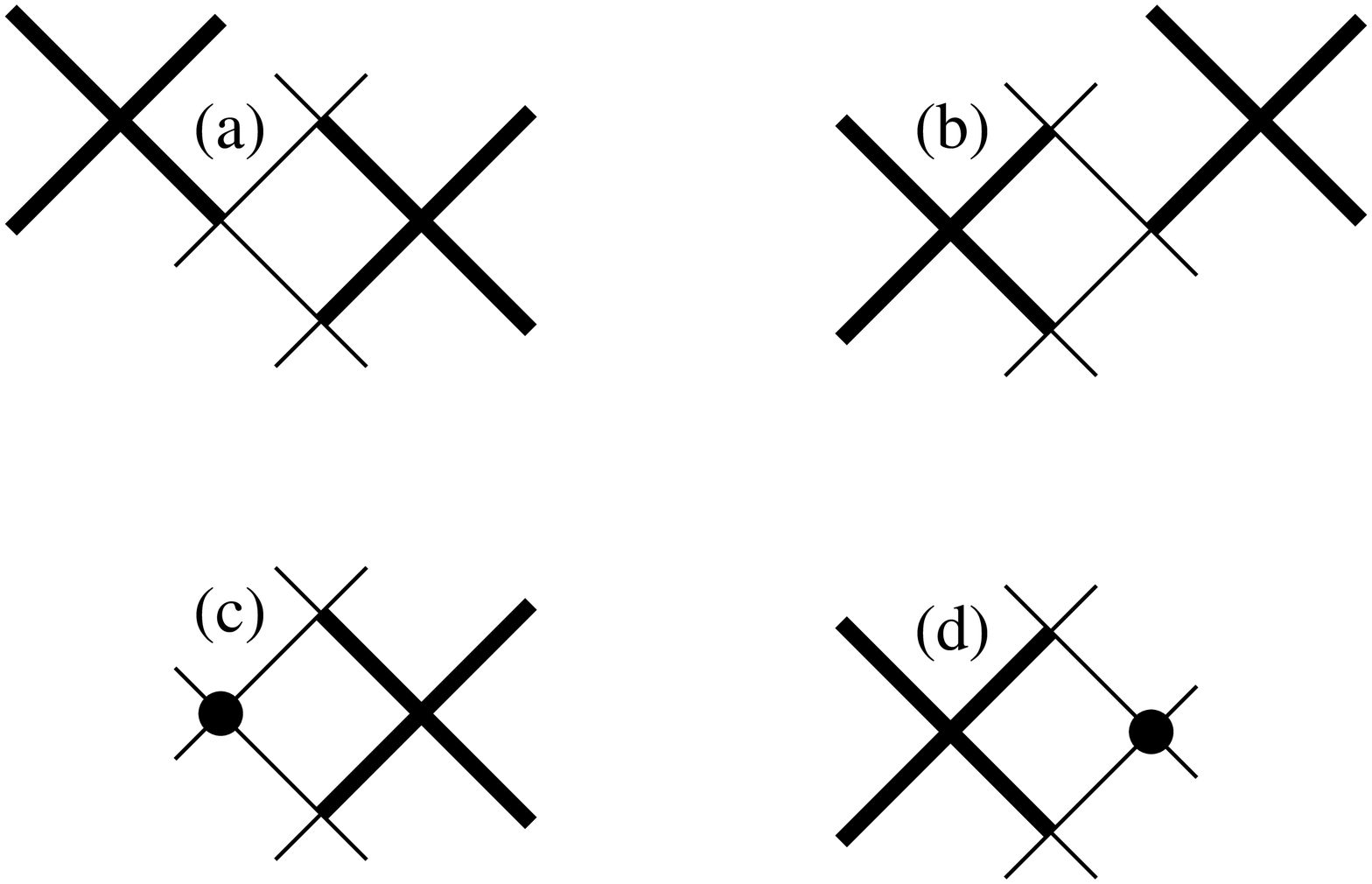,width=3in}
\end{center}
\par
FIG. 18 \vspace{5cm}
\end{figure}

\newpage

\begin{figure}[p]
\begin{center}
\epsfig{file=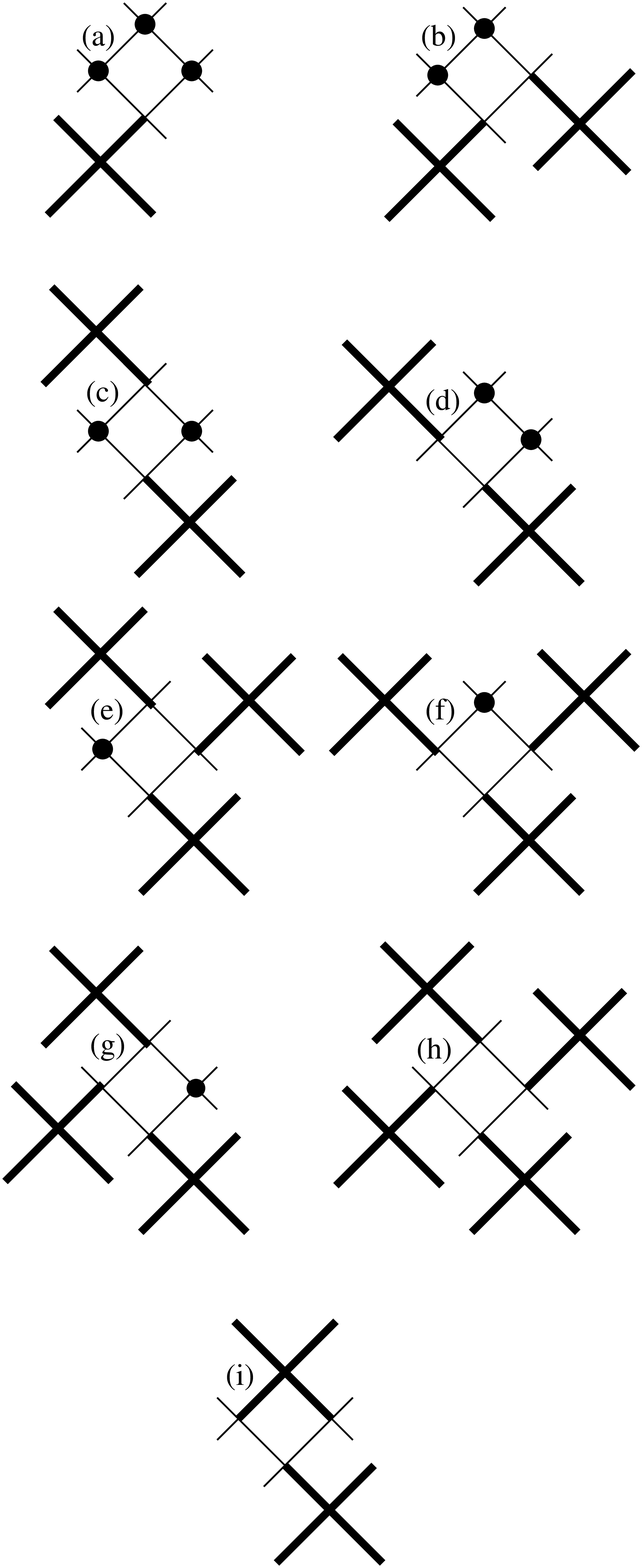,width=3in}
\end{center}
\par
FIG. 19 \vspace{5cm}
\end{figure}

\newpage

\begin{figure}[p]
\begin{center}
\epsfig{file=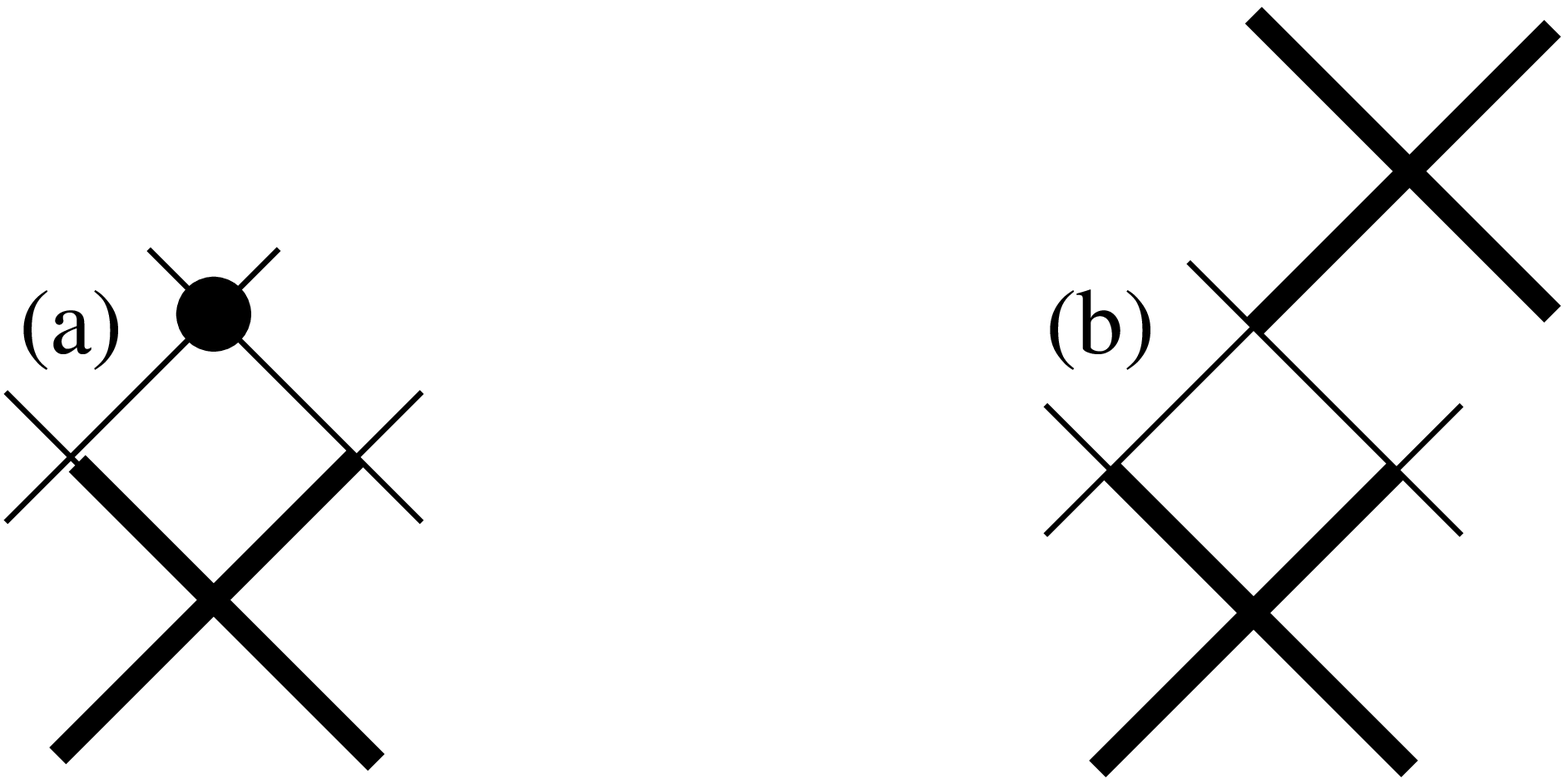,width=3in}
\end{center}
\par
FIG. 20 \vspace{5cm}
\end{figure}

\newpage

\begin{figure}[p]
\begin{center}
\epsfig{file=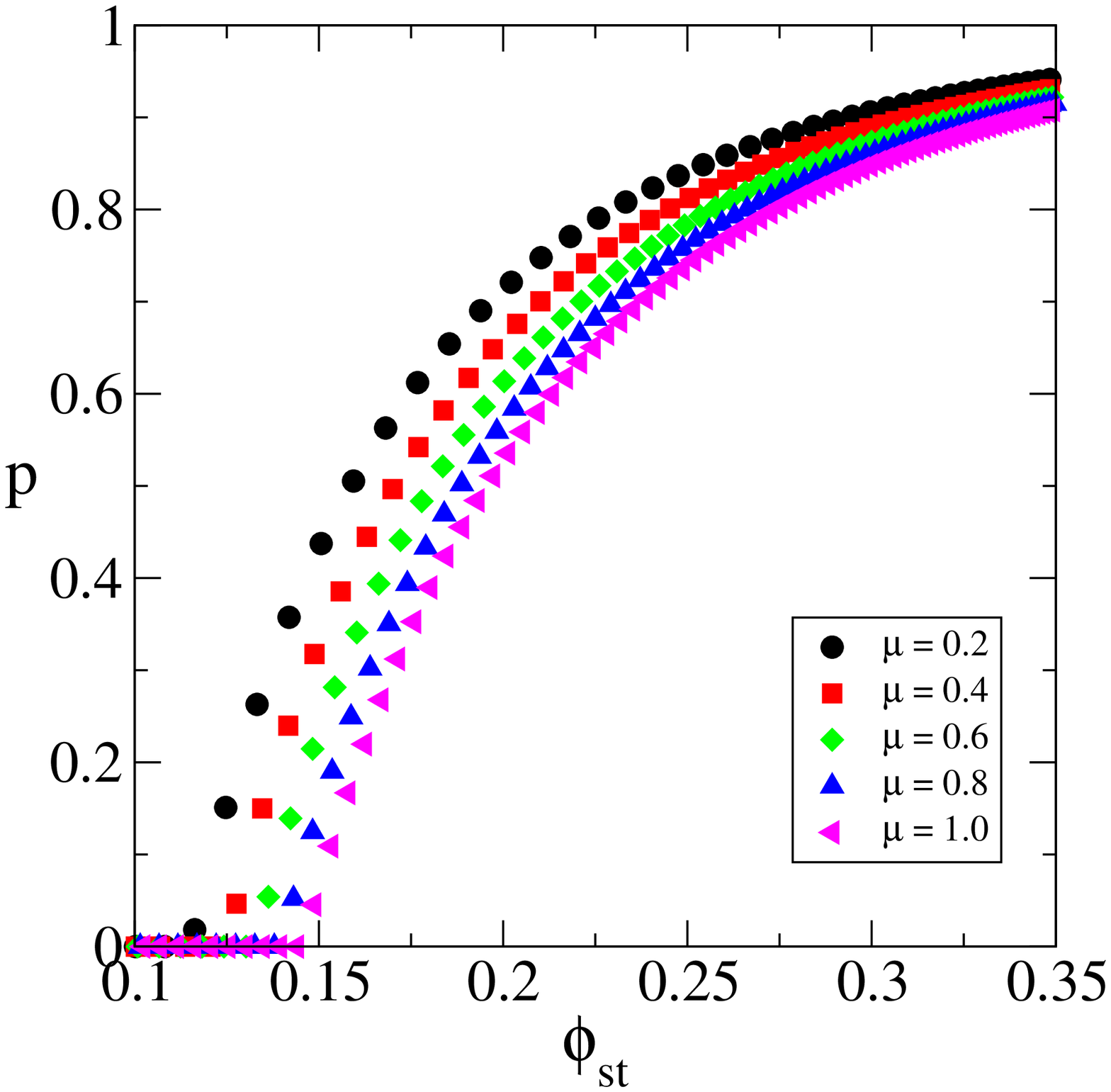,width=8.5cm,angle=270}
\end{center}
\par
FIG. 21 \vspace{5cm}
\end{figure}

\newpage

\begin{figure}[p]
\begin{center}
\epsfig{file=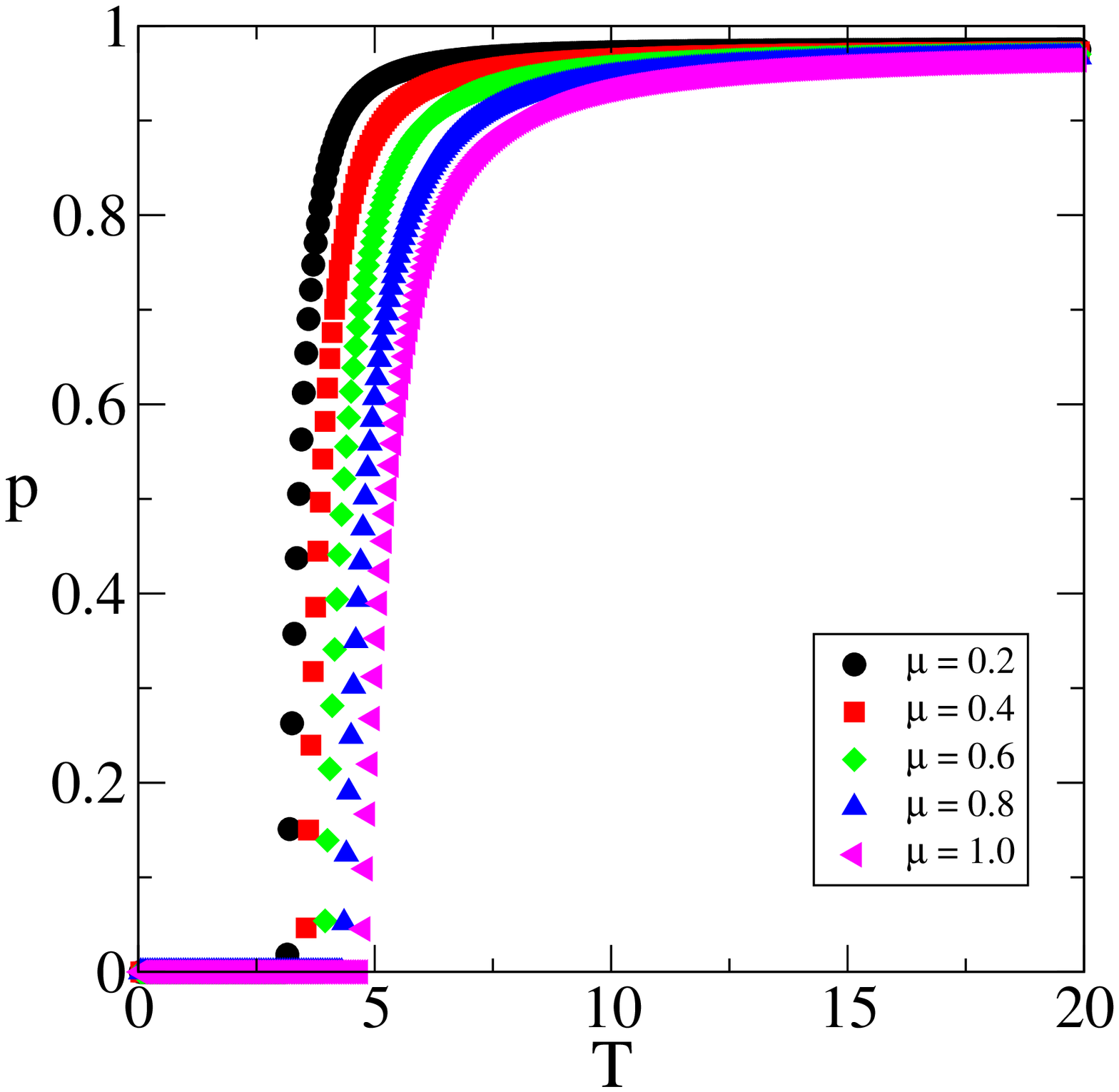,width=8.5cm,angle=270}
\end{center}
\par
FIG. 22 \vspace{5cm}
\end{figure}

\newpage

\begin{figure}[p]
\begin{center}
\epsfig{file=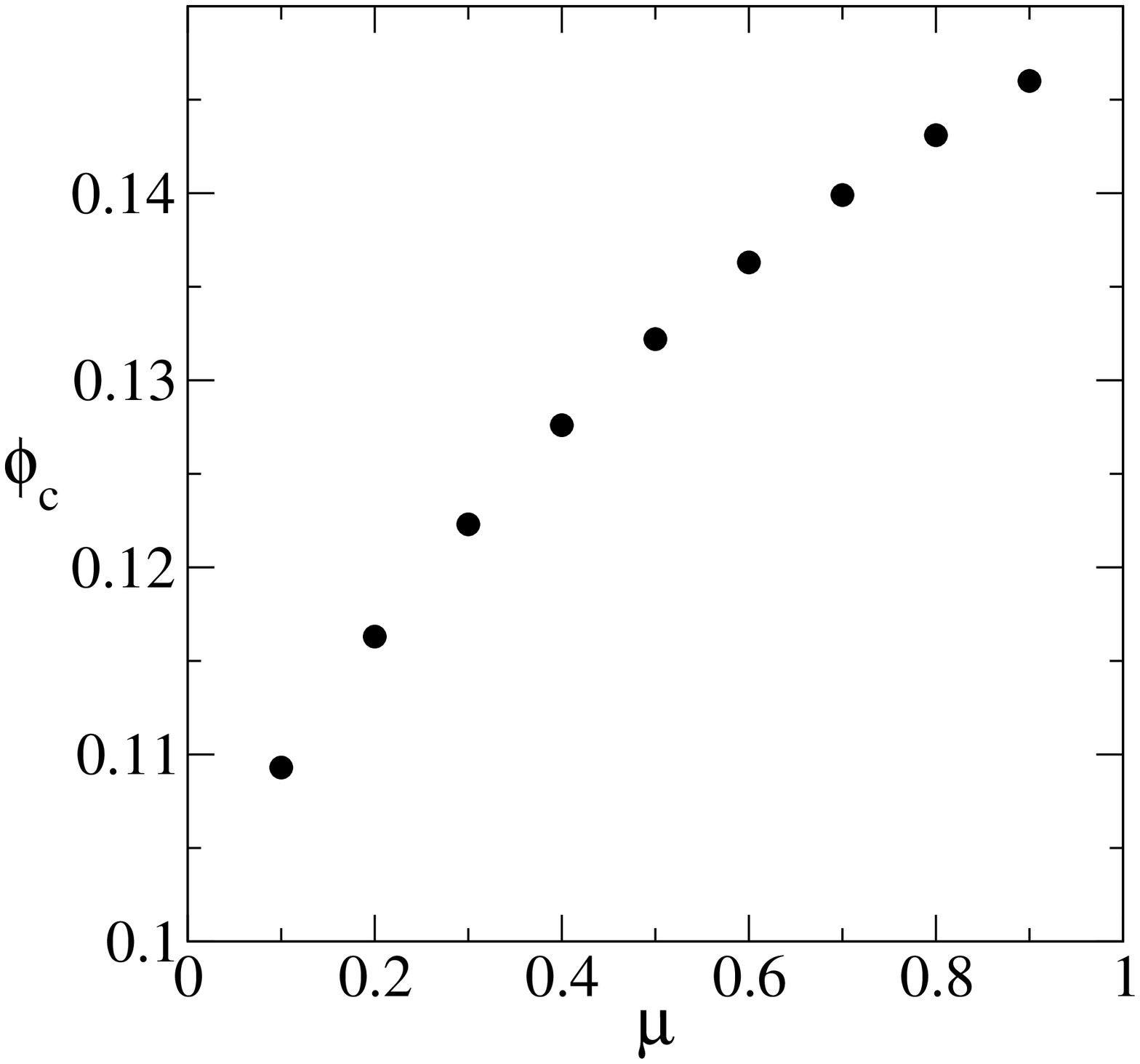,width=8.5cm,angle=270}
\end{center}
\par
FIG. 23 \vspace{5cm}
\end{figure}

\newpage

\begin{figure}[p]
\begin{center}
\epsfig{file=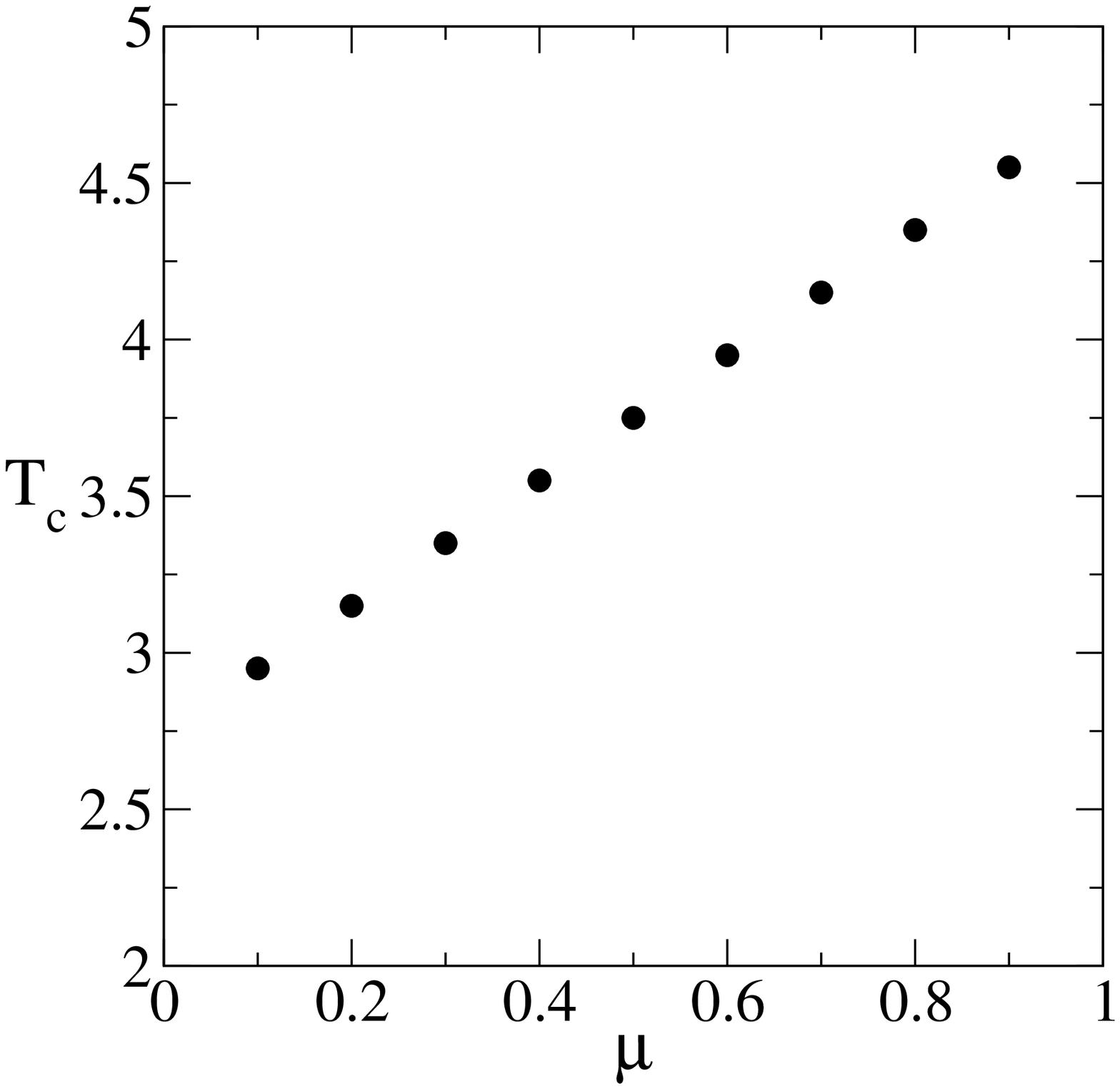,width=8.5cm,angle=270}
\end{center}
\par
FIG. 24 \vspace{5cm}
\end{figure}

\clearpage

\newpage

\begin{figure}[p]
\begin{center}
\epsfig{file=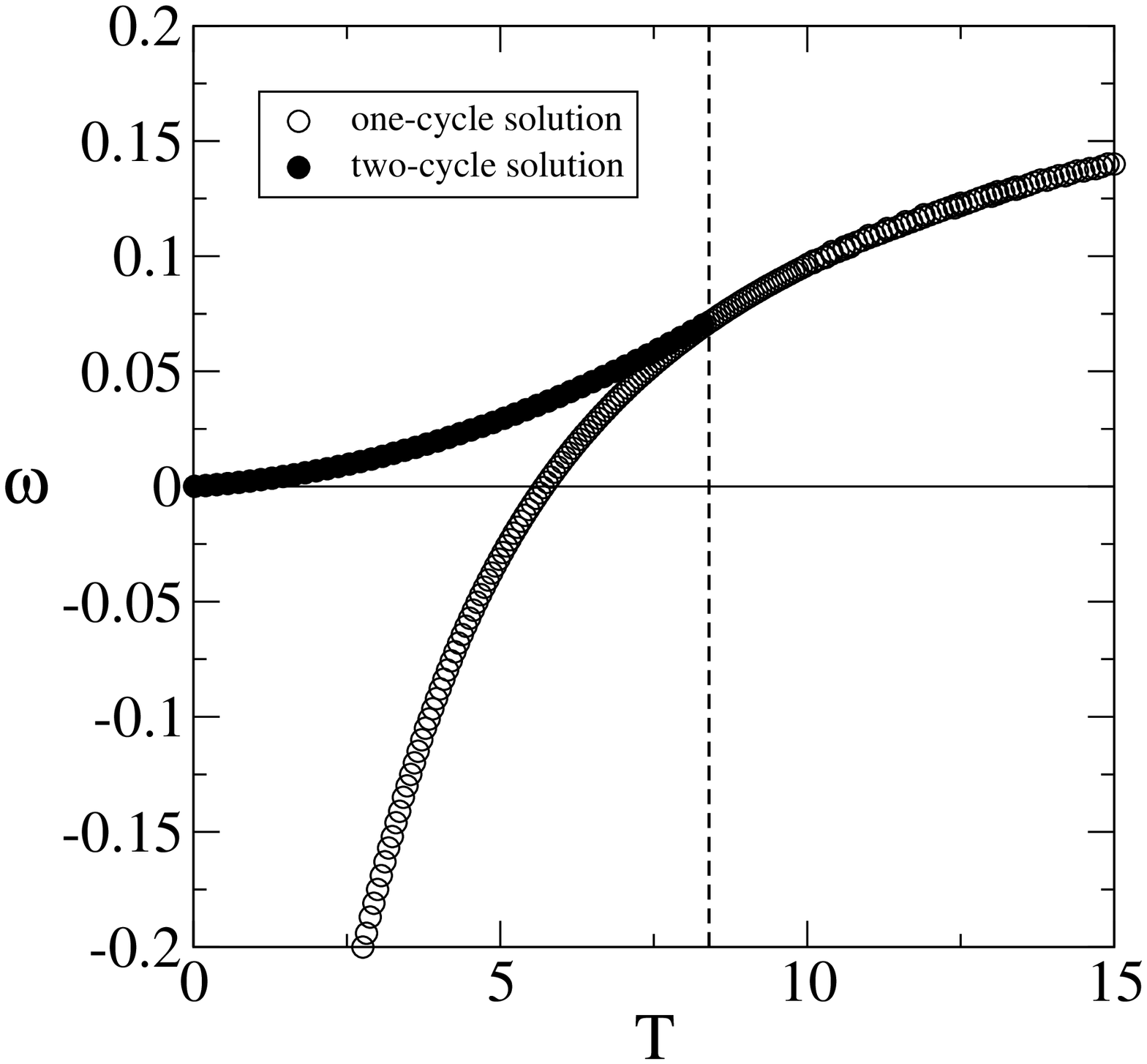,width=8.5cm,angle=270}
\end{center}
\par
FIG. 25 \vspace{5cm}
\end{figure}

\newpage

\begin{figure}[p]
\begin{center}
\epsfig{file=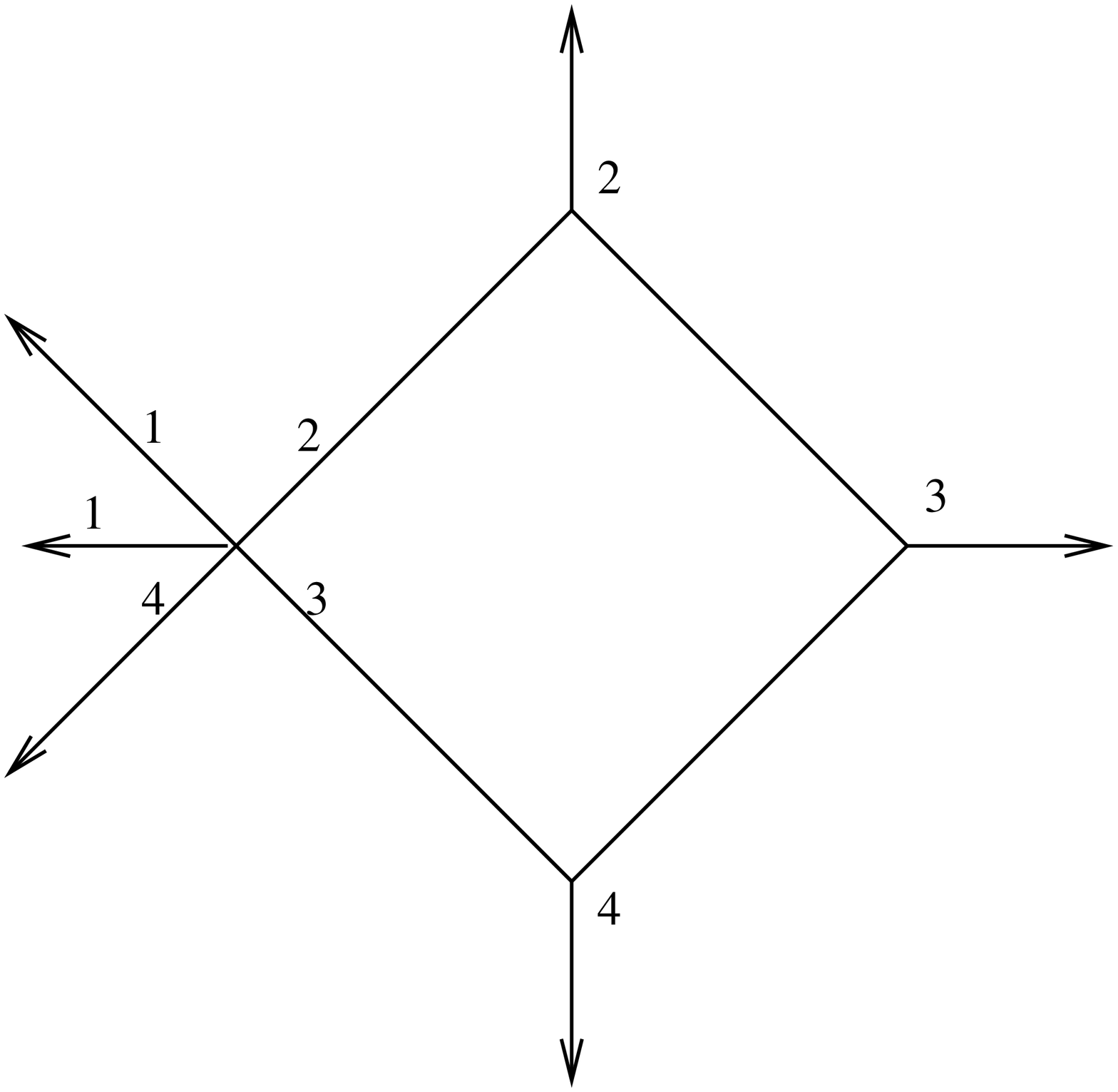,width=8.5cm}
\end{center}
\par
FIG. 26 \vspace{5cm}
\end{figure}

\newpage

\begin{figure}[p]
\begin{center}
\epsfig{file=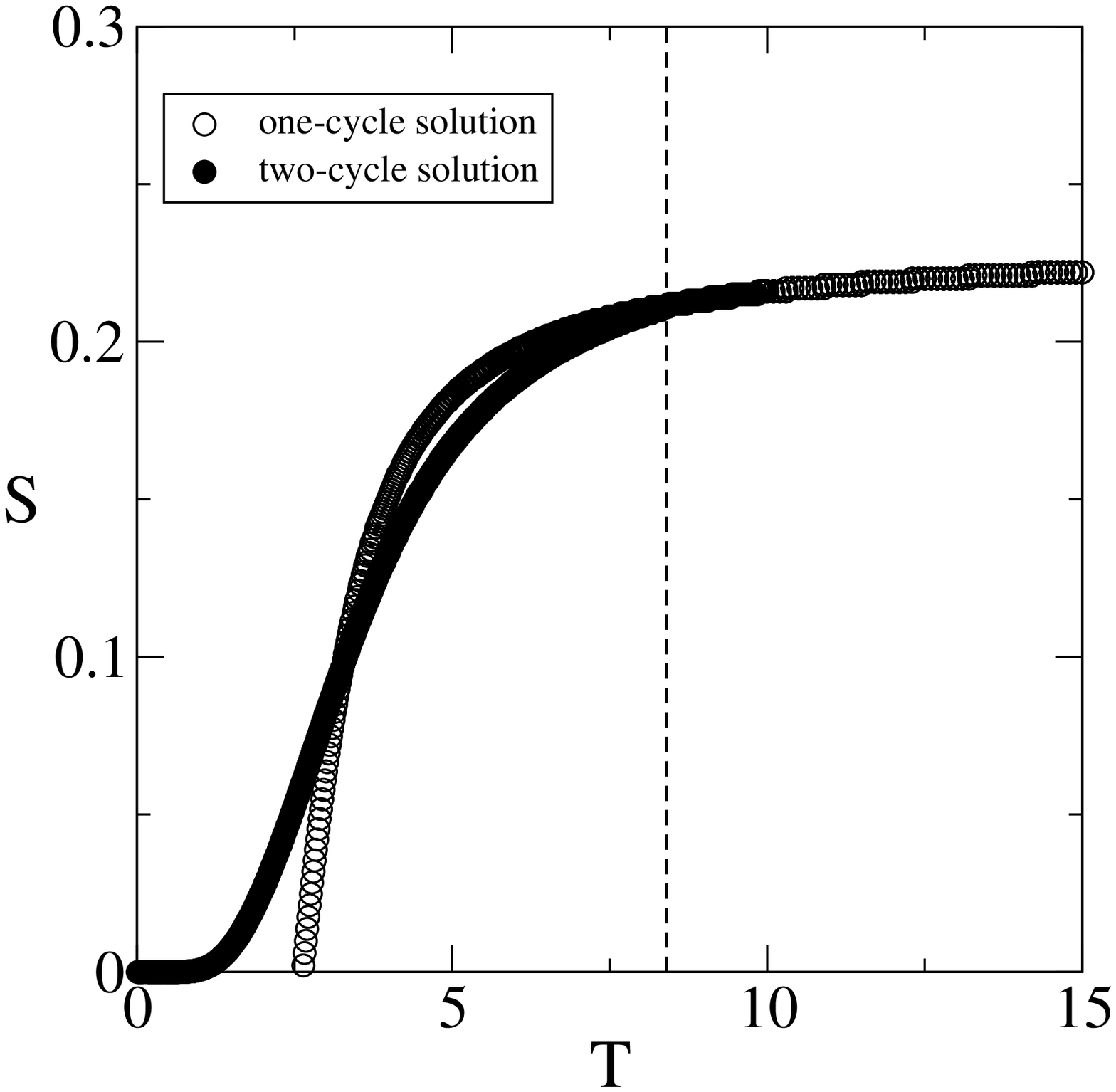,width=8.5cm,angle=270}
\end{center}
\par
FIG. 27 \vspace{5cm}
\end{figure}

\newpage

\begin{figure}[p]
\begin{center}
\epsfig{file=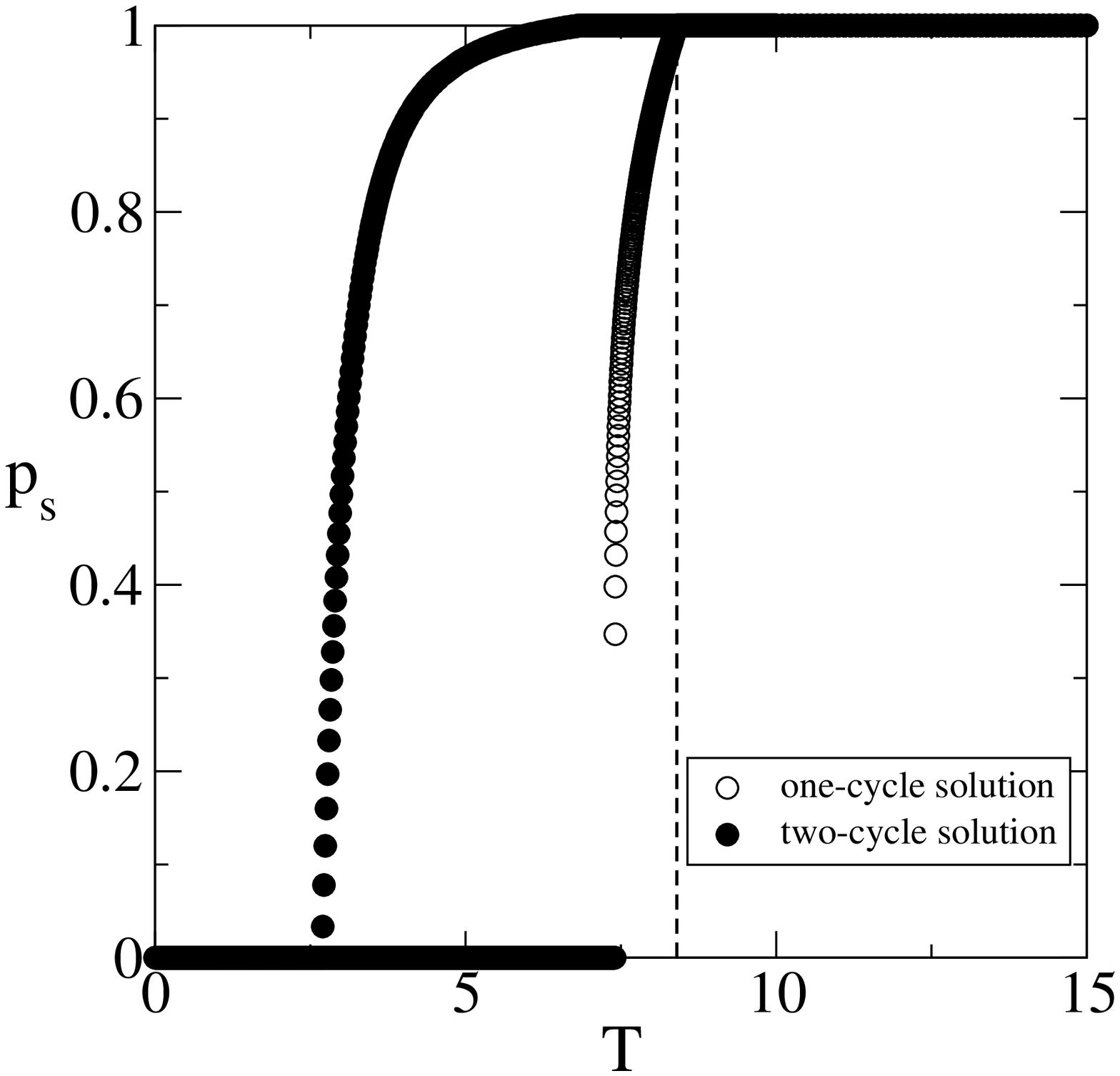,width=8.5cm,angle=270}
\end{center}
\par
FIG. 28 \vspace{5cm}
\end{figure}

\newpage

\begin{figure}[p]
\begin{center}
\epsfig{file=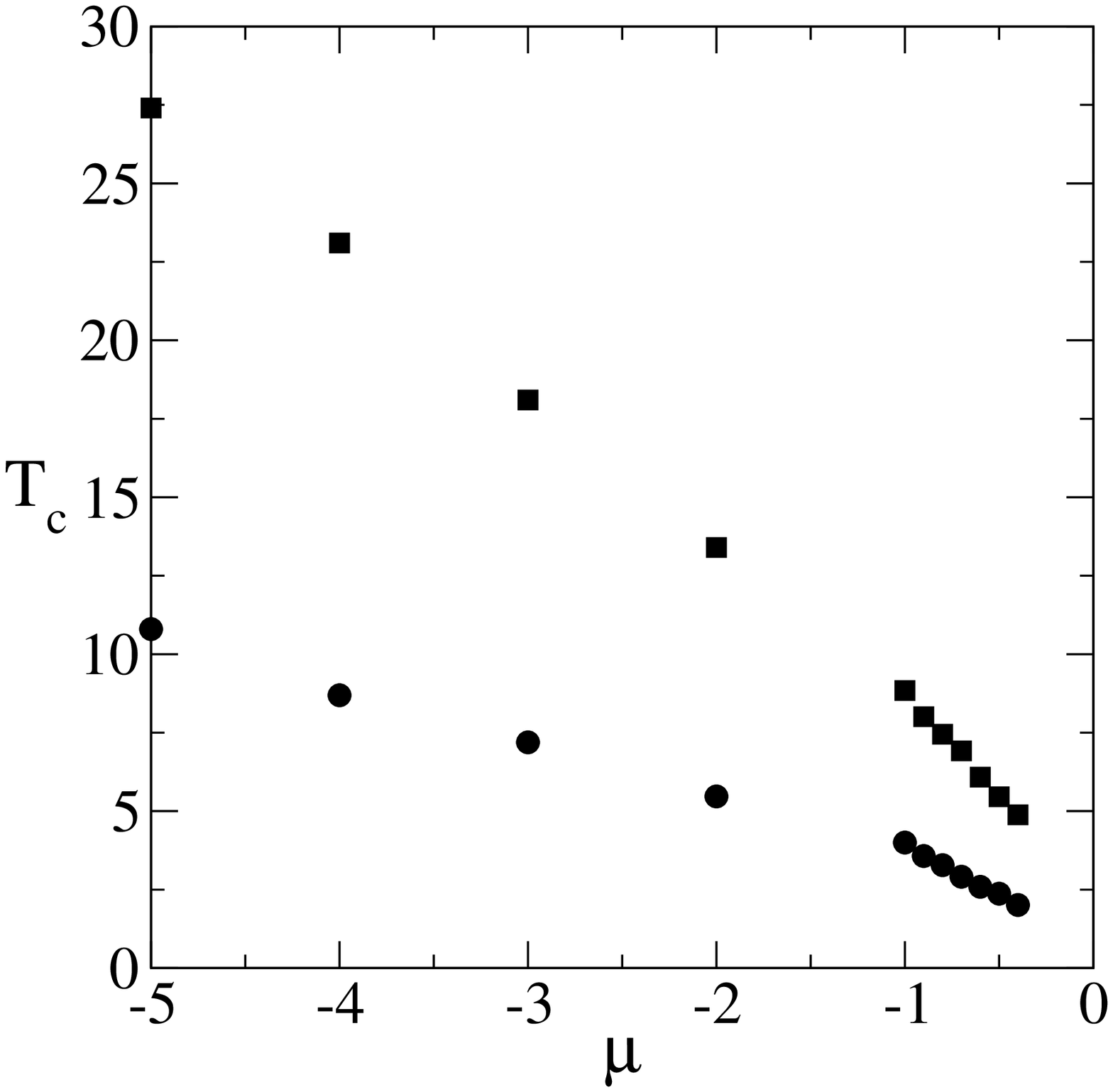,width=8.5cm,angle=270}
\end{center}
\par
FIG. 29 \vspace{5cm}
\end{figure}

\end{document}